\documentclass[11pt]{article}
\usepackage{graphicx}
\usepackage{epstopdf}
\usepackage{amsfonts,amssymb}
\usepackage{amsmath,amsthm}
\usepackage{bbm,enumerate}
\usepackage{algorithm,algorithmic}
\usepackage{graphicx}
\usepackage{graphicx}
\usepackage{epsf}
\usepackage{mathrsfs}
\usepackage{subfigure}
\usepackage{epsfig}
\usepackage{latexsym}
\usepackage{graphics}
\usepackage{color,hyperref}
\newtheorem{thm}{Theorem}
\newtheorem{thm-alpha}{Theorem~\ref{thm:alpha-rate-inapprox}}
\newtheorem{cor}{Corollary}
\newtheorem{prop}{Proposition}

\newtheorem{lem}{Lemma}

\newcommand{\nop}[1]{}

\newcommand{\remove}[1]{}

\author{Ferdinando Cicalese\thanks{Dipartimento di Informatica, University of Salerno, 84084 Fisciano (SA), Italy. Email: \texttt{cicalese@dia.unisa.it}.}
\and Martin Milani\v c
\thanks{University of Primorska, UP IAM, Muzejski trg 2, SI6000 Koper, Slovenia, and
University of Primorska, UP FAMNIT, Glagolja\v ska 8, SI6000 Koper, Slovenia. Email: \texttt{martin.milanic@upr.si}.}
\and
Ugo Vaccaro\thanks{Dipartimento di Informatica, University of Salerno, 84084 Fisciano (SA), Italy. Email: \texttt{uv@dia.unisa.it}.}}
\date{}

\title{On the approximability and exact algorithms for vector domination and related problems in graphs}

\begin{document}
\maketitle

\begin{abstract}
We consider two graph optimization problems called vector domination and total vector domination.
In vector domination one seeks a small subset $S$ of vertices of a graph  such that
any vertex outside $S$ has a prescribed number of neighbors in $S$.
In total vector domination, the requirement is extended to all vertices of the graph.
We prove that these problems (and several variants thereof) cannot be approximated to within a factor of $c\ln n$,
where $c$ is a suitable constant and $n$ is the number of the vertices,
unless {\sf P} = {\sf NP}.
We also show that two natural greedy strategies have approximation factors
$\ln \Delta+O(1)$, where $\Delta$ is the maximum degree of the input graph.
We also provide exact polynomial time algorithms for several classes of graphs.
Our results extend, improve,  and  unify several results previously known
in the literature.
\end{abstract}



\section{Introduction }

The concept of domination in graphs has been extensively
studied, both in structural and algorithmic graph theory,
because of  its numerous  applications to a variety of areas.
Informally, a set of vertices of a graph is said to dominate
a vertex if it contains a sufficient part of its closed neighborhood, where the exact definition of ``sufficient''
depends on the model.
Generally, one seeks small sets
that dominate the whole graph.
Domination naturally arises in facility location problems,
in problems involving finding sets of representatives, in
monitoring communication or electrical networks, and in land surveying.
The two books \cite{HHS1-98} \cite{HHS2-98} discuss the main results and
applications of domination in graphs.
Many variants of the basic concepts of domination
have appeared in the literature. Again, we refer to
\cite{HHS1-98} \cite{HHS2-98} for a survey of the area.

In this paper we provide  hardness results, approximation algorithms and exact polynomial
time algorithms for an interesting generalization of the basic
concept of domination, firstly introduced in \cite{HPV-99}.
Here, a subset of vertices $S$ is said to dominate
a vertex $v$ if either $v\in S$, or  there are in $S$
a \emph{prescribed} number of neighbors of $v$ (see below
for formal definitions). Again, one seeks small subsets
that dominate (in this new sense) the whole vertex set of the graph.


\smallskip

\noindent
{\bf Main Definitions.}
For a graph $G=(V,E)$ and a vertex $v\in V$,
denote by $N(v)$ (or $N_G(v)$, if the graph is not clear from the context)
the set of neighbors of $v$ (that is, the {\it (open) neighborhood} of $v$),
by $N[v]:= N(v)\cup \{v\}$ (or $N_G[v]$) the {\it closed neighborhood} of $v$, by $d(v)=d_G(v)$
the degree of $v$, and by $\Delta(G)$  the maximum degree of any vertex in $G$.
A {\em dominating set} in a graph $G=(V,E)$ is a subset of the graph's vertex set such that every vertex not in the set has a neighbor in the set. A {\em total dominating set} in $G$ is a subset $S\subseteq V$ such that every vertex of the graph has a neighbor in $S$, that is, for every $v\in V$ there exists a vertex $u\in S$ such that $uv\in E$.
\begin{sloppypar}
The {\it vector domination} is the following problem: Given a graph $G = (V,E)$, and
a vector ${\bf k} = (k_v:v\in V)$ such that for all $v\in V$, $k_v\in \{0,1,\ldots, d(v)\}$,
find a {\it vector dominating set} (VDS) of minimum size, that is, a set $S\subseteq V$ minimizing $|S|$
and such that $|S\cap N(v)|\ge k_v$ for all $v\in V\backslash S$.
Vector dominating sets were introduced in \cite{HPV-99}, have also appeared in the literature under the name of {\em threshold ordinary dominating sets}~\cite{GH-07}, and were recently studied from the parameterized complexity point of view~\cite{RSS}.
The {\it total vector domination} is the problem of finding a minimum-sized {\it total vector dominating set}, that is, a set $S\subseteq V$ such that $|S\cap N(v)|\ge k_v$ for all $v\in V$.
The minimum sizes of vector and total vector dominating sets will be denoted
by $\gamma(G,{\bf k})$ and $\gamma^t(G,{\bf k})$, respectively.
If in the definition of total vector domination we replace open neighborhoods with closed ones, we get the so called {\it multiple domination} problem~\cite{LC-02,LC-03}: Given a graph $G=(V,E)$ and a vector ${\bf k} = (k_v:v\in V)$ such that for all $v\in V$, $k_v\in \{0,1,\ldots, d(v)+1\}$, find a minimum size set $S\subseteq V$ such that for all vertices $v\in V$, it holds that $|N[v]\cap S|\ge k_v$.
We will also consider the following
special cases of vector domination, total vector domination, and multiple domination:
\begin{itemize}
  \item For $0<\alpha\le 1$, an {\em $\alpha$-dominating set} in $G$ is a subset $S\subseteq V$ such that every vertex not in the set has at least an $\alpha$-fraction of its neighbors in the set, that is, for every $v\in V\setminus S$, it holds that $|N(v)\cap S|\ge  \alpha|N(v)|$.
  \item For $0<\alpha\le 1$, a {\em total $\alpha$-dominating set} in $G$ is
a subset $S\subseteq V$ such that every vertex has at least an $\alpha$-fraction of its neighbors in the set, that is, for every $v\in V$, it holds that $|N(v)\cap S|\ge  \alpha|N(v)|$.
  \item For $0<\alpha\le 1$, an {\em $\alpha$-rate dominating set} in $G$ is a subset $S\subseteq V$ such that every vertex has at least an $\alpha$-fraction of the members of its closed neighborhood in the set, that is, for every $v\in V$, it holds that $|N[v]\cap S|\ge  \alpha|N[v]|$.
\end{itemize}
By $\gamma(G)$ ($\gamma_\alpha(G)$,
$\gamma^t(G)$, $\gamma^t_\alpha(G), \gamma_{\times \alpha}(G)$) we denote the minimum size of a dominating ($\alpha$-dominating, total dominating, total $\alpha$-dominating, $\alpha$-rate dominating) set in $G$.
For a fixed $0 <\alpha \le 1$, the problem of finding in a given graph a dominating (\hbox{$\alpha$-dominating}, total dominating, total $\alpha$-dominating, $\alpha$-rate dominating) set of minimum size will be referred to simply as the {\it domination} (\hbox{$\alpha$-domination}, {\it total domination}, {\it total $\alpha$-domination}, {\it $\alpha$-rate domination}). The notion of $\alpha$-domination was introduced by Dunbar {\em et al.}~\cite{DHLM-2000} and studied further in~\cite{DRV-2004,DRV-2008,GPZ-09,GPZ-11}. The notion of $\alpha$-rate domination was introduced
in 2009 by Gagarin {\em et al.}~\cite{GPZ-09,GPZ-11}. To the best of our knowledge, the notions of
total vector domination and total $\alpha$-domination are new, that is, they are introduced
in~\cite{CMV-11} and in this paper, respectively.
\end{sloppypar}

Notice that for every $\alpha>0$, every $\alpha$-dominating set is a
dominating set,  every total $\alpha$-dominating set is a
total dominating set, every vertex cover is an $\alpha$-dominating set,
and every $1$-dominating set is a vertex cover.
As shown in~\cite{DHLM-2000}, for graphs of maximum vertex degree at most $1/\alpha$,  $\alpha$-dominating sets coincide with dominating sets, and it can be shown similarly that
for such graphs total $\alpha$-dominating sets coincide with total dominating sets.
Moreover, for graphs of maximum vertex degree less than $1/(1-\alpha)$ (where $\alpha <1$),
$\alpha$-dominating sets coincide with vertex covers~\cite{DHLM-2000}.
Clearly, the (total) $\alpha$-domination corresponds to the  special case of the (total) vector domination,
in which $k_v = \lceil\alpha\cdot d(v)\rceil$ for all $v\in V$.
In fact, we shall mainly use $\alpha$-domination for our inapproximability results,
and provide algorithmic results in terms of the more general problem of vector domination.

{In Table~\ref{fig:table} we summarize definitions of several domination parameters.\footnote{
The notions of  {\it strict $\alpha$-domination} and {\it strict total $\alpha$-domination}
were introduced under different names in~\cite{CMV-11}.
}

{
\begin{table}[ht]
  \footnotesize\centering
    \begin{tabular}{|l|c|c|c|c|}
  \hline
  Model & Neighborhood & Total / Partial & Inequal. & Threshold type\\
  \hline
  $\alpha$-domination~\cite{DHLM-2000} &  open &  partial & $\ge$ & fraction $= \alpha$\\
  \hline
  $\alpha$-rate domination~\cite{GPZ-09} & closed &  total & $\ge$ & fraction $= \alpha$\\
  \hline
  domination~\cite{HHS1-98} & closed &  total & $\ge$ & uniform, $k_v = 1$ $\forall v$ \\
  \hline
  $k$-domination~\cite{FJ-85a,FJ-85b} & open &  partial & $\ge$ & uniform, $k_v = k$ $\forall v$ \\
  \hline
  $k$-tuple domination~\cite{HH00} & closed &  total & $\ge$ & uniform, $k_v = k$ $\forall v$ \\
  \hline
  $k$-tuple total domination~\cite{HK-10} & open &  total & $\ge$ & uniform, $k_v = k$ $\forall v$ \\
  \hline
  monopoly~\cite{Peleg-02} &  closed &  total & $\ge$ & fraction $= 1/2$ \\
  \hline
  multiple domination~\cite{LC-02,LC-03} & closed &  total & $\ge$ & non-uniform\\
\hline
  partial monopoly~\cite{Peleg-02} &  open &  partial & $>$ & fraction $= 1/2$ \\
 \hline
  positive influence domination~\cite{Wang11} &  open  &  total & $\ge$ & fraction $= 1/2$  \\
 \hline
  strict $\alpha$-domination~\cite{CMV-11} &  open  &  partial & $>$ & fraction $= \alpha$  \\
 \hline
  strict total $\alpha$-domination~\cite{CMV-11} &  open  &  total & $>$ & fraction $= \alpha$  \\
   \hline
  total $\alpha$-domination [this paper] &  open &  total & $\ge$ & fraction $= \alpha$\\
\hline
  total domination~\cite{HHS1-98} &  open &  total & $\ge$ & uniform, $k_v = 1$ $\forall v$\\
\hline
  total vector domination~\cite{CMV-11} &  open &  total & $\ge$ & non-uniform \\
  \hline
  vector domination~\cite{HPV-99} &  open &  partial & $\ge$ & non-uniform \\
  \hline
  vertex cover~\cite{HHS1-98} & open &  partial & $\ge$ & $k_v = d(v)$ $\forall v$ \\
  \hline
\end{tabular}
\caption{Definitions of different domination models. The neighborhoods can be either {\it open} $(N(v))$ or {\it closed} ($N[v]$); the domination constraint can be either
required for all $v\in V$ ({\it total}) or only for $v\in V\setminus S$ ({\it partial}); the type of inequality can be
either weak ($\ge$) or strict ($>$); the threshold can be either {\it uniform} ($k_v = k$ for all $v\in V$),
{\it non-uniform} (every $v\in V$ has its own threshold $k_v$) or expressed as a {\it fraction} of the size of the (open or closed)
neighborhood ($\alpha\cdot N_v$, where $N_v\in \{N(v),N[v]\}$), according to the neighborhood type as specified in the second column.
Notice that some of the models can be defined by more than one choice of the parameters.}
\label{fig:table}
\end{table}
}

\noindent
{\bf Our Results.}
We first provide  two natural greedy algorithms  for vector domination and
total vector domination in general graphs, having
approximation factors of $\ln(2\Delta(G))+1$ and $\ln({\Delta(G)})+1$, respectively.
Subsequently, we prove that the above results are essentially best
possible, in the sense that both the $\alpha$-domination and its total variant are
inapproximable within a factor of $c\ln n$ for a suitable constant $c>0$, unless
${\sf P} = {\sf NP}$. We also obtain better inapproximability bounds under the stronger hypothesis ${\sf NP}\nsubseteq {\sf DTIME}(n^{O(\log\log n)})$.
Notice that our inapproximability results are provided for {\em any} fixed $0 <\alpha < 1$, and this range of values for $\alpha$
is as large as it can be: the $0$-domination, $0$-total domination, and $1$-total domination problems are trivial,
and the $1$-domination problem coincides with the $2$-approximable vertex cover problem.
We also obtain inapproximability results for other problems listed in Table~\ref{fig:table} (with the exception of vertex cover),
see Section~$3$ and the summarizing Table~\ref{fig:table2} in Section~$5$.


 Subsequently, we individuate special classes of graphs for which vector domination and total vector domination can be optimally solved in polynomial time.
 More specifically, we provide polynomial time algorithms for
 computing minimum size vector domination sets and total
 vector domination sets for complete graphs, trees, $P_4$-free graphs and threshold graphs.


\noindent{\bf Related Work.}
The papers~\cite{DRV-2004,DHLM-2000,GPZ-09} provide several bounds for the value of $\gamma_\alpha(G)$
in terms of other graph parameters, while~\cite{DRV-2008} gives
a characterization of $\alpha$-perfect trees.
{In addition, Dunbar {\em et al.}~\cite{DHLM-2000} give an {\sf NP}-completeness
result for $\alpha$-domination. This result is extended considerably by our
inapproximability results. Conversely, our approximability result for vector domination
answers an open problem posed in~\cite{GPZ-09,GPZ-11} where the authors suggest to develop
algorithms approximating $\alpha$-domination to a certain degree of precision.}

The algorithmic aspect of total vector domination in strongly
chordal graphs (a super-class of trees) was studied in \cite{LC-03},
where a polynomial time algorithm for that purpose was given. However,
the authors of \cite{LC-03} point out that their approach cannot
be modified to handle the case of vector domination, and that a new approach
is needed.


   {Strictly related to our results is also the paper \cite{MRS-2002}.
 The authors study the hardness of approximating minimum monopolies in graphs~\cite{Peleg-02}.
%
In the language of Table~\ref{fig:table}, a monopoly corresponds to a $(1/2)$-rate dominating set,
and a partial monopoly to a strict $(1/2)$-dominating set. Therefore,  our inapproximability
results for $\alpha$-rate domination and strict $\alpha$-domination can be seen as
significant extensions of the results of \cite{MRS-2002} from the case $\alpha=1/2$ to arbitrary $\alpha$.}
It is also worth mentioning a recent paper on the approximability of the majority monopoly problem~\cite{M-2012}.



\begin{sloppypar}
 Our findings are also  relevant to
 the new area of influence spread
 in social networks \cite{KKT-05}, specifically, to
 positive influence dominating sets (PIDS) in social networks~\cite{Wang11}.
 In our language, PIDS correspond to
 total $\alpha$-dominating sets with $\alpha = 1/2$.
%
In \cite{Wang11} it is proved that PIDS is APX-hard.
Our hardness results for total $\alpha$-domination are
 more general, and also stronger since  we prove
 inapproximability within a logarithmic factor.
In the same area, the paper \cite{ZWZW-10}
 introduced the problem of identifying a minimum set of
 nodes that could influence a whole network within a time bound $d$.
 There, a set of nodes $S$ influences a new node $x$ in one step  ($d=1$) if
 the majority  of neighbors of $x$ is in  $S$. The paper  \cite{ZWZW-10}
 mostly studies hardness results for the case $d=1$.
 It is clear that our scenario includes that of \cite{ZWZW-10} (in the case $d=1$)
 and corresponds to a more extensive model of influence among nodes, similar to the one considered in
 \cite{MR-07} for a related but different problem.
\end{sloppypar}




\section{Approximability results}\label{sec:approx}

{In this section, we show that  vector domination and total vector domination can be approximated in polynomial time by a factor of $\ln(2\Delta(G))+1$ and $\ln(\Delta(G))+1$, respectively. (We denote by $\ln$ the natural logarithm.)

We start with total vector domination and related problems. Our results are based on the results for the
set cover problem. Consider the following generalization of the set cover problem:

\noindent\texttt{SET MULTICOVER}

\noindent{\it Instance:} A set-system ${\cal C} = (U,{\cal F})$, where $U$ is a finite ground set
and ${\cal F}$ is a collection of subsets of $U$;
a non-negative integer requirement ${\it req}(u)$ for every element $u$ of the ground set.

\noindent{\it Task:}  Find a minimum size subcollection ${\cal F}'\subseteq {\cal F}$ such that
every element $u$ appears in at least ${\it req}(u)$ sets in ${\cal F}'$.

\medskip
The decision version of the \texttt{SET MULTICOVER}
problem is {\sf NP}-complete~\cite{GJ79}. Moreover,
the greedy algorithm produces a solution that is always within a factor $\ln \Delta+1$ of the optimum,
where $\Delta$ is the maximum size of a set in ${\cal F}$~\cite{Dobson}.

\bigskip
Every instance of any of the ``total''~domination problems defined in Table~\ref{fig:table}
(see the third column of the table)
can be described as an instance of the \texttt{SET MULTICOVER} problem.
For example, if $(G,{\bf k})$ is an instance to the total vector domination problem,
we can take $U = V(G)$, define ${\cal F}$ to be the collection of all (open) neighborhoods, and set ${\it req}(u) = k_u$ for all $u\in U$.
It is clear that a subset $S\subseteq V(G)$ is a total vector dominating set for $(G,{\bf k})$ if and only if the collection
$(N(v):v\in S)$ is a feasible solution to the instance $(U,{\cal F}, {\it req})$ of the \texttt{SET MULTICOVER} problem.
Similar transformations work for the other ``total"~domination problems.

We therefore obtain the following results and their corollaries:

\begin{thm}\label{thm:vtdom-approx}
Total vector domination can be approximated in polynomial time by a factor of $\ln(\Delta(G))+1$.
\end{thm}

\begin{cor}
Total $\alpha$-domination,  strict total $\alpha$-domination, $k$-tuple total domination, and
positive influence domination problems can be approximated in polynomial time by a factor of
$\ln(\Delta(G))+1$.
\end{cor}

\begin{thm}
The multiple domination problem can be approximated in polynomial time by a factor of $\ln(\Delta(G)+1)+1$.
\end{thm}

\begin{sloppypar}
\begin{cor}
The $\alpha$-rate domination, the $k$-tuple domination problem and the monopoly problem can be approximated in polynomial time by a factor of $\ln(\Delta(G)+1)+1$.
\end{cor}
\end{sloppypar}

The $(\ln(\Delta(G))+1)$-approximability of the positive influence domination problem and
the $(\ln(\Delta(G)+1)+1)$-approximability of the $k$-tuple domination problem
were proved in~\cite{Wang11} and~\cite{KL04}, respectively.

\bigskip
The above approach does not seem to be easily applicable to ``partial''~domination problems
such as vector domination, $\alpha$-domination, $k$-domination, partial monopoly and strict $\alpha$-domination.
Instead, we will show below that these problems can be recast as a particular case of the well known \texttt{MINIMUM SUBMODULAR COVER} problem, and apply a classical result due to Wolsey~\cite{Wolsey-82}.}

\begin{thm}\label{thm:vdom-approx}
Vector domination
can be approximated in polynomial time by a factor of {$\ln({2\Delta(G)})+1$}.
\end{thm}

\begin{proof}
For a graph $G=(V,E)$ and a vector ${\bf k} = (k_v:v\in V)$ s.t.  for all $v\in V$, $k_v\in \{0,1,\ldots, d(v)\}$, we define
a  function
$f:2^V\longrightarrow \mathbb{N}$, as follows:
\begin{equation}\label{eq:potential:1}
\begin{array}{c}
\mbox{for all $S\subseteq V$}\,, \quad \mbox{let}  \quad \displaystyle{f(S) = \sum_{v\in V}\tau_v(S)}\,,
\quad \mbox{where }\\
\\
\tau_v(S) = \left\{
           \begin{array}{ll}
             \min\{|S\cap N(v)|, k_v\}, & \hbox{if $v\not\in S$;} \\
             k_v, & \hbox{if $v\in S$.}
           \end{array}
         \right.
\end{array}
\end{equation}
The following properties of $f$ can be verified: (i) $f$ is integer valued; (ii) $f(\emptyset) = 0;$
(iii) {\it $f$ is non-decreasing};
(iv) {\it A set  $S\subseteq V$ satisfies $f(S) = f(V)$ if and only if $S$ is a vector dominating set};
(v) {\it $f$ is submodular.}

Recall that a function $f:2^V\longrightarrow\mathbb{N}$ is \emph{submodular}
if for all $S\subseteq T\subseteq V$ and for all $w\in V\setminus T$,
the inequality $f(T\cup \{w\})-f(T)\le f(S\cup \{w\})-f(S)$ holds.
The only non-trivial property to show is (v), i.e, the submodularity of  $f$.
The  proof  is given  below.

\begin{lem}\label{lem:subm2}
The function $f:2^V\longrightarrow \mathbb{N}$, given by~(\ref{eq:potential:1}), is submodular.
\end{lem}

\begin{proof}
%
It suffices to show that all the functions $\tau_v(\cdot)$ are submodular, that is, that for all $S\subseteq T\subseteq V$ and for all $w\in V\setminus T$,
\begin{equation}\label{eq:subm}
\tau_v(T\cup \{w\})-\tau_v(T)\le \tau_v(S\cup \{w\})-\tau_v(S)\,.
\end{equation}
Observe that $\tau_v$ is non-decreasing.
%
%

Suppose first that $\tau_v(T) = k_v$. Then $\tau_v(T\cup \{w\}) = k_v$ and
the left-hand side of inequality~(\ref{eq:subm}) is equal to 0. Hence
inequality~(\ref{eq:subm}) holds since $\tau_v$ is non-decreasing.

From now on, we assume that $\tau_v(T) <k_v$, which implies $\tau_v(T)= |T\cap N_G(v)|$.
Since $\tau_v$ is non-decreasing, $\tau_v(S)<k_v$, and hence
$\tau_v(S)= |S\cap N_G(v)|$.
Inequality~(\ref{eq:subm}) simplifies to
\begin{equation}\label{eq:subm2}
\tau_v(T)-\tau_v(S) = |(T\setminus S)\cap N_G(v)| \ge \tau_v(T\cup \{w\})-\tau_v(S\cup \{w\})\,.
\end{equation}
We may assume that $\tau_v(T\cup \{w\})>\tau_v(S\cup \{w\})$,
since otherwise the right-hand side
of~(\ref{eq:subm2}) equals~0, and inequality~(\ref{eq:subm2}) holds.

Therefore, $\tau_v(S\cup \{w\})<k_v$, implying $\tau_v(S\cup \{w\}) = |(S\cup \{w\})\cap N_G(v)|$.
If also $\tau_v(T\cup \{w\}) <k_v$ then
$\tau_v(T\cup \{w\}) = |(T\cup \{w\})\cap N_G(v)|$ and equality holds in (\ref{eq:subm2}).

So we may assume that $\tau_v(T\cup \{w\}) = k_v$. Note that $v$ does not belong to
$T\cup \{w\}$ (since otherwise either $\tau_v(T)$ or $\tau_v(S\cup \{w\})$ would equal to~$k_v$).
Suppose that the inequality~(\ref{eq:subm2}) fails. Then
$$|(T\setminus S)\cap N_G(v)|<k_v-|(S\cup\{w\})\cap N_G(v)|\,,$$
which implies
$$|(T\cup \{w\})\cap N_G(v)|<k_v\,.$$
However, together with the fact that $v\not\in T\cup\{w\}$, this contradicts the
assumption that $\tau_v(T\cup \{w\}) = k_v$.

\end{proof}

Back to the proof of Theorem \ref{thm:vdom-approx}, by (iv) we have that an optimal solution to the
 vector dominating set is provided by a minimum size $S$ such that $f(S) = f(V).$
In other words, we have recast vector domination as a particular case of
the \texttt{MINIMUM SUBMODULAR COVER} \cite{Wolsey-82}.

Let ${\mathbb A}$ denote the  natural greedy strategy which starts with  $S = \emptyset$ and
iteratively adds to $S$ the element $v \in V\setminus S$ s.t.
$f(S \cup \{v\}) - f(S)$ is maximum, until $f(S) = f(V)$ is achieved.
By a classical result of Wolsey~\cite{Wolsey-82}, it follows that algorithm $\mathbb{A}$
is a $(\ln(\max_{y \in V} f(\{y\}))+1)$-approximation algorithm
for vector domination.
For every $y\in V$, we have
$f(\{y\}) = \sum_{v\in V\setminus\{y\}}\tau_v(\{y\}) + \tau_y(\{y\}) \le d(y)+ k_y \le 2d(y).$
Hence $\max_{y \in V} f(\{y\})\le 2\Delta(G)$ yielding the desired result.
\end{proof}

{Since $\alpha$-domination, $k$-domination, partial monopoly and strict $\alpha$-domination problems
are all special cases of the vector domination problems, Theorem~\ref{thm:vdom-approx} implies the following result:
\begin{cor}
\begin{sloppypar}$\alpha$-domination, $k$-domination, partial monopoly and strict $\alpha$-domination problems
can be approximated in polynomial time by a factor of $\ln({2\Delta(G)})+1$.
\end{sloppypar}\end{cor}}

\section{Inapproximability results}\label{sec:inapprox}

Recall the following result on the inapproximability of domination and total domination,
which was derived from the analogous result about the set cover problem due to Feige~\cite{Feige-98}.
Hereafter, $n$ denotes the number of vertices of the input graph.

\begin{sloppypar}
\begin{thm}\cite{CC04}\label{thm:D-inapprox}
For every $\epsilon >0$, there is no polynomial time algorithm approximating
domination (total domination) for graphs without isolated vertices within a factor of \hbox{$(1-\epsilon)\ln n$}, unless
${\sf NP}\subseteq {\sf DTIME}(n^{O(\log\log n)})$.
\end{thm}
\end{sloppypar}

Most of our inapproximability results are given in terms of the variants of the $\alpha$-domination problem.
In fact, it turns out that $\alpha$-domination, its total variant, and the $\alpha$-rate domination are
inapproximable within a $c\ln n$ factor (for suitable constants $c$) as shown in
Theorems \ref{thm:qdom-inapprox}, \ref{thm:qtdom-inapprox} and \ref{thm:alpha-rate-inapprox} below.
{\em A fortiori}  the same results hold for the vector domination, total vector domination and
multiple domination problems.  Hence, the approximation results of the previous section are essentially best possible.
We shall  use the following lemma which is an {\em ad hoc} extension of the hardness of approximating  domination within
$(1-\epsilon)\ln n$ given in \cite{CC04}.

{\begin{lem} \label{lem:dom-set-inapprox}
For every integer $B>0$ and for every $\epsilon>0$,
there is no polynomial time algorithm approximating
domination on input graphs $G$ without isolated vertices satisfying $\gamma(G)\ge B\Delta(G)$
within a factor of \hbox{$(\frac{1}{2}-\epsilon)\ln n$}, unless
${\sf NP}\subseteq {\sf DTIME}(n^{O(\log\log n)})$.
\end{lem}

\begin{proof}
Let $B$ be a positive integer and $\epsilon\in (0,\frac{1}{2})$. We make a reduction from domination
on graphs without isolated vertices (Theorem~\ref{thm:D-inapprox}).
Let $G$ be a graph without isolated vertices with $|V(G)|\ge B^{1/\epsilon}$ that is an instance to domination.
We transform $G$ into a graph $G'$ which consists of
$N = B\Delta(G)$ disjoint copies of $G$, say $G_1,\ldots, G_{N}$.
Then clearly $\gamma(G') = N\gamma(G)$, while $\Delta(G') = \Delta(G)$.
In particular, since $\gamma(G)\ge 1$, the graph $G'$ satisfies
$\gamma(G')\ge N=B\Delta(G')$.

\begin{sloppypar}
For brevity, let us write $n = |V(G)|$ and $n' = |V(G')|$.
Suppose that there exists a polynomial time algorithm $A'$ that computes a
$(\frac{1}{2}-\epsilon)\ln n'$-approximation to domination in $G'$.
Let $S'$ be the set computed by $A'$.
Then $|S'|\le (\frac{1}{2}-\epsilon)(\ln n') \gamma(G')$.
\end{sloppypar}

For $i = 1,\ldots, N$, let $S_i'= S'\cap V(G_i)$, and let $S=S_{i^*}'$ such that
$|S_{i^*}'|\le |S_{i}'|$ for all $1\le i \le N$.
Then $S$ is a dominating set in (the $i^*$-th copy of) $G$.
Moreover, we can bound the size of $S$ from above as follows:
$$
\begin{array}{rcll}
|S| &\le& (1/N)\cdot |S'| & \textrm{(by the definition of $S$)}\\
& \le & (1/N)\cdot (\frac{1}{2}-\epsilon)(\ln n')\cdot \gamma(G')& \textrm{(by the assumption on $A'$)}\\
& = & (\frac{1}{2}-\epsilon)\ln(B\Delta(G)n)\cdot\gamma(G) & \textrm{(by the properties of $G'$)}\\
& \le & (\frac{1}{2}-\epsilon)\ln(n^{2+\epsilon})\cdot\gamma(G) & \textrm{(since $B\le n^{\epsilon}$ and $\Delta(G)\le n$)}\\
& = & (\frac{1}{2}-\epsilon)(2+\epsilon)(\ln n)\cdot\gamma(G) &\\
& = & (1- \epsilon')(\ln n)\cdot\gamma(G)\,, &\\
\end{array}$$
where $\epsilon' := 3/2\epsilon +\epsilon^2\in (0,1)$.
Therefore, if there exists a polynomial time
algorithm that computes a
$(\frac{1}{2}-\epsilon)\ln n'$-approximation to domination in $G'$,
there exists a polynomial time algorithm that computes a
$(1-\epsilon')\ln n$-approximation to domination in $G$, and hence, by
Theorem~\ref{thm:D-inapprox}, this is only possible if ${\sf NP}\subseteq {\sf DTIME}(n^{O(\log\log n)})$.
\end{proof}

\begin{thm}\label{thm:qdom-inapprox}
For every $\alpha\in (0,1)$ and every $\epsilon>0$,
there is no polynomial time algorithm approximating
$\alpha$-domination within a factor of \hbox{$(\frac{1}{2}-\epsilon)\ln n$}, unless
${\sf NP}\subseteq {\sf DTIME}(n^{O(\log\log n)})$.
\end{thm}

\begin{proof}
Let $0<\alpha<1$ and $\epsilon\in (0,\frac{1}{2})$.
We define $N = \lceil\frac{\alpha}{1-\alpha}\rceil$ and $B = \lceil N/ \epsilon\rceil$.
Let $G$ be a graph without isolated vertices, with $\gamma(G)\ge B\Delta(G)$ and such that
$|V(G)|\ge(N+1)^{1/\epsilon}$. We transform $G$ into a graph $G'$ which consists of $G$ together with a
set $K$ of $k = N\Delta(G)$ vertices such that $K$ is disjoint from $V(G)$. In addition, every vertex $v$ from $G$
is adjacent to precisely $k_v$ vertices in $K$, where
$$k_v = \left\{
          \begin{array}{ll}
            \left\lceil\frac{\alpha d_G(v)-1}{1-\alpha}\right\rceil, & \hbox{if $d_G(v)\ge 2$;} \\
            0, & \hbox{if $d_G(v) = 1$.}
          \end{array}
        \right.$$
(This assignment is done in an arbitrary way.)
Notice that the $k_v$'s are well defined, since $G$ has no isolated vertices.
Moreover, if $d_G(v) = 1$ then $k_v= 0<k$, while for $d_G(v)\ge 2$ we have
$$0\le k_v
= \left\lceil\frac{\alpha d_G(v)-1}{1-\alpha}\right\rceil
< \frac{\alpha d_G(v)-1}{1-\alpha} +1
<\frac{\alpha}{1-\alpha}d_G(v) \le
N\Delta(G) = k\,.$$
Hence it is indeed possible to assign to $v$ precisely $k_v$ neighbors in $K$.

In addition, $k_v$ is an integer satisfying
$$\frac{k_v}{d_G(v)+k_v}<\alpha\le \frac{k_v+1}{d_G(v)+k_v}\,.$$
These inequalities are instrumental to the following result.

\noindent
{\em Claim:}
$\gamma(G)\le \gamma_\alpha(G') \le \gamma(G) + k\,.$

{\em Proof.}
Let $S'$ be an optimal $\alpha$-dominating set in $G'$.
Then, the set $S:=S'\cap V(G)$ is a dominating set in $G$. Indeed,
suppose for a contradiction that there exists a vertex $v$ in $G$ such that
$S$ misses the closed neighborhood of $v$.
Then $|N_{G'}(v)\cap S'|\le k_v$. The degree of $v$ in $G'$ is equal to
$d_{G'}(v) = d_G(v)+k_v$.
Therefore
$$\frac{|N_{G'}(v)\cap S'|}{d_{G'}(v)}\le \frac{k_v}{d_G(v)+k_v}<\alpha\,,$$
contrary to the assumption that $S'$ is $\alpha$-dominating.
Consequently $\gamma(G) \le |S'\cap V(G)|\le |S'| = \gamma_\alpha(G')$.

Conversely, let $S$ be an optimal dominating set in $G$.
The set $S':= S\cup K$ is then an $\alpha$-dominating set in $G'$ such that
$|S'|= \gamma(G)+k$. To see that
$S'$ is $\alpha$-dominating in $G'$, observe that
for every $v\in V(G')\setminus S' = V(G)\backslash S$,
the set $N_{G'}(v)\cap S'$ is the disjoint union of sets
    $N_G(v)\cap S$ and $N_{G'}(v)\cap K$.
Hence
$$|N_{G'}(v)\cap S'| = |N_G(v)\cap S|+ |N_{G'}(v)\cap K|\ge 1 + k_v\ge\alpha(d_G(v)+k_v) = \alpha |N_{G'}(v)|\,,$$
        where the second inequality holds by the choice of $k_v$.
Altogether, this shows that $\gamma_\alpha(G') \le |S'| = \gamma(G) + k$ and completes the proof of the claim.

Again, let us write $n = |V(G)|$ and $n' = |V(G')|$.
Notice that we have
$N+1\le n^{\epsilon}$ and hence
$$n' = n+N\Delta(G) \le  n+Nn \le n^{1+\epsilon}\,.$$
Moreover,
$\frac{1}{\epsilon}k= \frac{1}{\epsilon}N\Delta(G)\le B\Delta(G)\le \gamma(G)$, which implies
$k\le \epsilon \gamma(G)$.

Suppose that there exists a polynomial time algorithm $A'$
which computes an $\alpha$-dominating set $S'$ for $G'$ such that
 $|S'|\le (\frac{1}{2}-\epsilon)(\ln n') \gamma_\alpha(G')$.
Let $S=S'\cap V(G)$. Like in the proof of the above claim, we see that $S$ is a dominating set in $G$.
We can bound the size of $S$ from above as follows:
$$
\begin{array}{rcll}
|S| &\le& |S'|&\\
& \le & (\frac{1}{2}-\epsilon)(\ln n')\cdot \gamma_\alpha(G')& \textrm{(by the assumption on $A'$)}\\
& \le & (\frac{1}{2}-\epsilon)(\ln n^{1+\epsilon})\cdot (\gamma(G)+k)& \textrm{(by the Claim and  $n'\le n^{1+\epsilon}$)}\\
& \le & (\frac{1}{2}-\epsilon)(1+\epsilon)(\ln n)\cdot (\gamma(G)+\epsilon\gamma(G))& \textrm{(since $k\le \epsilon\gamma(G)$)}\\
& = & (\frac{1}{2}-\epsilon)(1+\epsilon)^2(\ln n)\cdot\gamma(G)& \\%
& = & (\frac{1}{2}- \epsilon')(\ln n)\cdot\gamma(G)\,, &\\
\end{array}$$
where $\epsilon' := \epsilon^2(\epsilon+3/2)\in (0,\frac{1}{2})$.
%
Therefore, if there exists a polynomial time
algorithm that computes a $(\frac{1}{2}-\epsilon)\ln n'$-approximation to $\alpha$-domination in $G'$,
there exists a polynomial time algorithm that computes a
$(\frac{1}{2}-\epsilon')\ln n$-approximation to domination in $G$, and hence, by
Lemma \ref{lem:dom-set-inapprox}, this is only possible if
 ${\sf NP}\subseteq {\sf DTIME}(n^{O(\log\log n)})$.
\end{proof}}

{With minor modifications of the above proof (adapting it for strict inequalities, cf.~Table~\ref{fig:table}), one can obtain the analogue of Theorem~\ref{thm:qdom-inapprox} for {strict $\alpha$-domination}.

\begin{thm}\label{thm:strict-alpha-dom-inapprox}
For every $\alpha\in (0,1)$ and every $\epsilon>0$,
there is no polynomial time algorithm approximating
strict $\alpha$-domination within a factor of \hbox{$(\frac{1}{2}-\epsilon)\ln n$}, unless
${\sf NP}\subseteq {\sf DTIME}(n^{O(\log\log n)})$.
\end{thm}}

Theorem~\ref{thm:strict-alpha-dom-inapprox} is a significant extension of
the same result for the case $\alpha = 1/2$ proved in~\cite{MRS-2002}.
See also~\cite{CMV-11} for a proof of the inapproximability of strict $\alpha$-domination within a factor of the form $c\ln n$ for some constant $c>0$.

%
%

By means of a slightly more involved construction, we now prove a similar result for total $\alpha$-domination.

{
\begin{thm}\label{thm:qtdom-inapprox}
For every $\alpha\in (0,1)$ and every $\epsilon>0$,
there is no polynomial time algorithm approximating total
$\alpha$-domination within a factor of \hbox{$(\frac{1}{3}-\epsilon)\ln n$}, unless
${\sf NP}\subseteq {\sf DTIME}(n^{O(\log\log n)})$.
\end{thm}

\begin{proof}
Let $0<\alpha<1$ and $\epsilon\in (0,\frac{1}{3}).$
Let $B = \lceil\frac{\alpha}{1-\alpha}\rceil$.
We make a reduction from total domination on graphs $G$ with $n$ vertices such that
\begin{equation}\label{size-of-G-1}
n \ge \max\left\{\sqrt{\frac{1-\alpha}{\alpha}},2^{3/\epsilon},
(B+1)^{2/\epsilon}\right\}\,,
\end{equation}
\begin{equation}\label{size-of-G-2}
\lceil n^{1+\epsilon/3}\rceil\le n^{1+2\epsilon/3}\,,
\end{equation}
\begin{equation}\label{size-of-G-3}
n+B\le n^{1+\epsilon/3}\,.\end{equation}
and
\begin{equation}\label{gamma-t}
\gamma^t(G)\ge \frac{B}{\epsilon}\,.
\end{equation}
Clearly, these assumptions are without loss of generality since the inequalities in~(\ref{size-of-G-1})--(\ref{size-of-G-3}) are satisfied for all large enough $n$, while if the inequality (\ref{gamma-t}) is violated, we can find an optimal solution in polynomial time by verifying all subsets of $V(G)$ of size less than $\frac{B}{\epsilon}$.

Let $G$ be a graph satisfying $(\ref{size-of-G-1})$--$(\ref{gamma-t})$.
Let $n = |V(G)|$ and $m:=\lceil n^{1+\epsilon/3}\rceil$.
We transform $G$ into a graph $G'$ as follows:
$G'$ consists of $mn$ disjoint copies of $G$, say $G_1,\ldots, G_{mn}$,
together with a complete graph $K$ on $Bmn$ vertices such that
$K$ is disjoint from the $mn$ copies of $G$.
(See Fig.~\ref{fig:transformation2}.)
To describe the remaining edges,
we first partition the vertex set of $K$ into $m$ equally-sized parts $K_1,\ldots, K_{m}$.
(In particular, $|K_i| = Bn$ for all $i=1,\ldots, m$.) Finally, for every $j\in\{1,\ldots, mn\}$,
we make every vertex $v\in V(G_j)$ adjacent to precisely $k_v$ vertices in $K_{\lceil j/n\rceil}$
where $k_v$ is an integer satisfying
$$\frac{k_v}{d_G(v)+k_v}<\alpha\le\frac{1+k_v}{d_G(v)+k_v}\,.$$
Similarly as in the proof of Theorem~\ref{thm:qdom-inapprox},
we can take $k_v = \lceil\frac{\alpha d_G(v)-1}{1-\alpha}\rceil$ if $d_G(v)\ge 2$ and
$k_v = 0$ if $d_G(v) = 1$. (The graph $G$, as input to total domination, does not have any isolated vertices, since otherwise the problem is infeasible.)
Also notice that since $k_v\le Bn$, it is indeed possible to assign to every
$v\in V(G_j)$ precisely $k_v$ neighbors in $K_{\lceil j/n\rceil}$.
(This assignment is done in an arbitrary way.)

\begin{figure}[h!]
    \centering \includegraphics[width=120mm]{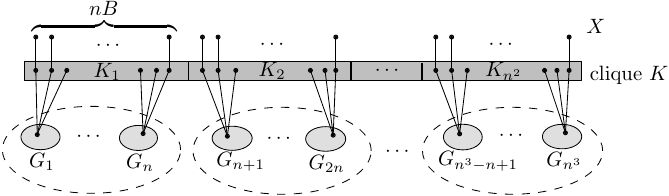}
\caption{The graph $G'$ in the proof of Theorem \ref{thm:qtdom-inapprox}}
\label{fig:transformation2}
\end{figure}

{\em Claim:} $mn\gamma^t(G)\le \gamma_\alpha^t(G') \le mn\gamma^t(G) + Bmn\,.$

\bigskip
Proof of Claim:

Let $S'$ be an optimal total $\alpha$-dominating set in $G'$, that is, $|S'| = \gamma_\alpha^t(G')$.
For every $j= 1,\ldots, mn$, let $S'_j = S'\cap V(G_j)$
denote the part of $S'$ that belongs to to the $j$-th copy of $G$ in $G'$.
Pick an index $j^*\in \{1,\ldots, mn\}$ for which the size of $S_j'$ is smallest.
We argue that the set $S:=S_{j^*}$ is a total dominating set
in $G_{j^*}$ (and thus in $G$). Indeed, suppose for contradiction that
there exists a vertex $v$ in $G_{j^*}$ such that $S$ misses the neighborhood of $v$.
Then $|N_{G'}(v)\cap S'|\le k_v$ while the degree of $v$ in $G'$ is equal to
$d_{G'}(v) = d_G(v)+k_v$.
Therefore $$\frac{|N_{G'}(v)\cap S'|}{d_{G'}(v)}\le \frac{k_v}{d_G(v)+k_v}<\alpha\,,$$
contrary to the assumption that $S'$ is total $\alpha$-dominating.
This implies that $\gamma^t(G)\le |S|$ and consequently
$mn\gamma^t(G)\le mn|S|\le \sum_{j = 1}^{mn}|S'_j|\le |S'| =\gamma_\alpha^t(G')$.


Conversely, let $S$ be an optimal total dominating set in $G$. For $j = 1,\ldots, mn$,
let $S_j$ denote the copy of $S$ in $G_j$, and let $S' = K\cup \bigcup_{j = 1}
^{mn}S_j$. The set $S'\subseteq V(G')$ satisfies $|S'|= mn\gamma^t(G) + Bnm$.
Moreover, $S'$ is a total $\alpha$-dominating set in $G'$:
\begin{itemize}
    \item For every $j= 1,\ldots, mn$ and for every $v\in V(G_j)$,
    the set $N_{G'}(v)\cap S'$ is the disjoint union of sets
    $N_{G_j}(v)\cap S_j$ and $N_{G'}(v)\cap K$.
Hence
$$|N_{G'}(v)\cap S'| =
|N_{G_j}(v)\cap S_j|+ |N_{G'}(v)\cap K|\ge$$$$\ge 1 + k_v\ge  \alpha(d_{G_j}(v)+k_v) = \alpha |N_{G'}(v)|\,.$$
        The second inequality holds by the choice of $k_v$.
    \item Let $v\in K$.
    By construction of $G'$, $v$ is adjacent to every other vertex in $K$, and to at most $n^2$ remaining vertices.
    Hence $d_{G'}(v)\le (|K|-1)+n^2=Bmn+n^2-1$. Moreover, $|N_{G'}(v)\cap S'|\ge |K|-1 = Bmn-1$.
    Therefore, to show that  $|N_{G'}(v)\cap S'|\ge \alpha |N_{G'}(v)|$, it suffices to prove that
    $$\frac{Bmn-1}{Bmn+n^2-1}\ge \alpha\,.$$
    This is equivalent to
    $Bmn(1-\alpha)\ge \alpha n^2-\alpha+1\,,$
    which (since $n^2\ge \frac{1-\alpha}{\alpha}$) follows from the inequality
    $Bmn(1-\alpha)\ge 2\alpha n^2\,,$
    or, equivalently,
    $$m\ge \frac{2\alpha}{(1-\alpha)B}n\,,$$
    which follows from the inequality
    $n^{\epsilon/3} \ge 2\,,$
    which, finally, holds true by~(\ref{size-of-G-1}).
\end{itemize}
This shows that $\gamma_\alpha^t(G') \le mn\gamma^t(G) + Bmn$
and completes the proof of the claim.

\medskip
%

Let us write $n' = |V(G')|$. By assumptions~(\ref{size-of-G-2}) and (\ref{size-of-G-3}) we have
$$n' = mn^2+Bmn \le
n^{3+2\epsilon/3} + Bn^{2+2\epsilon/3}
= (n+B)n^{2+2\epsilon/3}\le n^{1+\epsilon/3}n^{2+2\epsilon/3}= n^{3+\epsilon}\,.$$
Suppose that there exists a polynomial time algorithm $A'$ that computes a
$(\frac{1}{3}-\epsilon)\ln n'$-approximation to total $\alpha$-domination in $G'$.
Let $S'$ be the set computed by $A'$. Then $|S'|\le (\frac{1}{3}-\epsilon)(\ln n') \gamma^t_\alpha(G')$.

Similarly as in the proof of the claim above, let $S'_j = S'\cap V(G_j)$
and pick an index $j^*\in \{1,\ldots, mn\}$ for which the value of $|S_j'|$ is smallest.
Then, setting $S=S'_{j^*}$ results in a total dominating set in $G_j$ (and hence in $G$).

\begin{sloppypar}
We can bound the size of $S$ from above as follows:
$$
\begin{array}{rcll}
|S| &\le & \frac{1}{mn}\cdot|S'| & \textrm{(by the choice of $j^*$)}\\
& \le & \frac{1}{mn}(\frac{1}{3}-\epsilon)(\ln n') \cdot\gamma^t_\alpha(G')& \textrm{(by the assumption on $A'$)}\\
& \le & \frac{1}{mn}(\frac{1}{3}-\epsilon)(\ln(n^{3+\epsilon}))\cdot(mn\gamma^t(G) + Bnm)& \textrm{(by the Claim and $n' \le n^{3+\epsilon}$)}\\
& = & (\frac{1}{3}-\epsilon)({3+\epsilon})(\ln n)\cdot(\gamma^t(G) + B)& \\
& \le & (\frac{1}{3}-\epsilon)({3+\epsilon})(1+\epsilon)(\ln n)\cdot \gamma^t(G)& \textrm{(by (\ref{gamma-t}))}\\
& = & (1-\epsilon')(\ln n)\gamma^t(G)\,, &
\end{array}$$
where $\epsilon' = \epsilon^3+11\epsilon^2/3+5\epsilon/3\in (0,1)$.
Therefore, $S$ approximates the total domination within a factor of
$(1-\epsilon')\ln n$. By Theorem~\ref{thm:D-inapprox}, this shows that there is no polynomial time algorithm approximating
total $\alpha$-domination within a factor of \hbox{$(\frac{1}{3}-\epsilon)\ln n$}, unless
${\sf NP}\subseteq {\sf DTIME}(n^{O(\log\log n)})$.
\end{sloppypar}
\end{proof}}

{Again, with obvious modifications of the above proof, one can obtain the analogue of Theorem~\ref{thm:qtdom-inapprox} for {strict total $\alpha$-domination}. 

\begin{thm}
\label{thm:strict-total-alpha-dom-inapprox}
For every $\alpha\in (0,1)$ and every $\epsilon>0$,
there is no polynomial time algorithm approximating
strict total $\alpha$-domination within a factor of \hbox{$(\frac{1}{3}-\epsilon)\ln n$}, unless
${\sf NP}\subseteq {\sf DTIME}(n^{O(\log\log n)})$.
\end{thm}}

{A different minor modification of the above proof (perform a reduction from domination instead of total domination, and
define $k_v = \lceil\frac{\alpha d_G(v)+\alpha-1}{1-\alpha}\rceil$ for all $v \in V(G)$) can be used to obtain the analogue of Theorem~\ref{thm:qtdom-inapprox} for {$\alpha$-rate domination}:
\begin{thm}\label{thm:alpha-rate-inapprox}
For every $\alpha\in (0,1)$ and every $\epsilon>0$,
there is no polynomial time algorithm approximating $\alpha$-rate domination
within a factor of \hbox{$(\frac{1}{3}-\epsilon)\ln n$}, unless
${\sf NP}\subseteq {\sf DTIME}(n^{O(\log\log n)})$.
\end{thm}

A proof of Theorem~\ref{thm:alpha-rate-inapprox} can be found in the appendix. Theorem~\ref{thm:alpha-rate-inapprox} is a significant extension of
the result for the case $\alpha = 1/2$ proved in~\cite{MRS-2002}.
At the same time, this result complements the $\ln(\Delta(G)+1)$-approximation  algorithm for $\alpha$-rate domination
outlined in Section~\ref{sec:approx}, providing an almost complete answer to the question about the approximability of
$\alpha$-rate domination posed by Gagarin {\em et al.}~in~\cite{GPZ-09,GPZ-11}.}

We continue with inapproximability results for $k$-domination and $k$-tuple total domination.
With a similar approach to the one by Klasing and Laforest showing the inapproximability of $k$-tuple domination~\cite{KL04},
we obtain the following result.

\begin{sloppypar}
\begin{thm}
\label{thm:k-domination-inapprox}
For every $k\ge 1$ and every $\epsilon>0$,  there is no polynomial time algorithm approximating
$k$-domination within a factor of \hbox{$(1-\epsilon)\ln n$}, unless
${\sf NP}\subseteq {\sf DTIME}(n^{O(\log\log n)})$.
\end{thm}
\end{sloppypar}

\begin{proof}
We make a reduction from domination on graphs $G$ with $n$ vertices such that
\begin{equation}\label{ll1}
n +k-1\le n^{1+\epsilon}
\end{equation}
and
\begin{equation}\label{gamma-k}
\gamma(G)\ge \frac{(k-1)(1+\epsilon)}{\epsilon^2}\,.
\end{equation}
Clearly, this assumption is without loss of generality since the inequality in~(\ref{ll1})
is satisfied for all large enough $n$, while if the inequality (\ref{gamma-k}) is violated, we can find an optimal solution in polynomial time by verifying all subsets of $V(G)$ of size less than $\frac{(k-1)(1+\epsilon)}{\epsilon^2}$.
Notice that inequality (\ref{gamma-k}) is equivalent to the following inequality:
\begin{equation}\label{gamma-k'}
\gamma(G) + k-1\le \frac{1+\epsilon+\epsilon^2}{1+\epsilon}\cdot \gamma(G)\,.
\end{equation}

Let $G$ be a graph satisfying $(\ref{ll1})$--$(\ref{gamma-k})$.
We transform $G$ into a graph $G'$ by adding to it a set $K$ of $k-1$ vertices
inducing a complete graph such that $V(G)$ and $K$ are disjoint,
and connecting every vertex in $G$ to every vertex in $K$.
Let $\gamma^{(k)}(G')$ denote the minimum cardinality of a $k$-dominating set in $G'$.
For every dominating set  $S$ in $G$, the set $S\cup K$ is a $k$-dominating set
in $G'$. Hence
\begin{equation}\label{gamma-k''}
\gamma^{(k)}(G')
\le \gamma(G) + k-1\,.
\end{equation}

Let $n' = |V(G')|$, and suppose that there exists a polynomial time algorithm $A'$ that computes a
$(1-\epsilon)\ln n'$-approximation to $k$-domination in $G'$. Let $S'$ be the set computed by $A'$.
Then $|S'|\le (1-\epsilon)(\ln n') \gamma^{(k)}(G')$. Moreover, the set $S:= S'\cap V(G)$ is a dominating set in $G$.

\begin{sloppypar}
We can bound the size of $S$ from above as follows:
$$\begin{array}{rcll}
|S| &\le & |S'|\\
& \le & (1-\epsilon)(\ln n') \gamma^{(k)}(G')& \textrm{(by the assumption on $A'$)}\\
& \le & (1-\epsilon)(\ln (n+k-1))(\gamma(G) + k-1)& \textrm{(by (\ref{gamma-k''}) and $n' = n+k-1$)}\\
& \le  &(1-\epsilon)\left(\ln (n^{1+\epsilon})\right)\left(\frac{1+\epsilon+\epsilon^2}{1+\epsilon}\gamma(G)\right)& \textrm{(by (\ref{ll1}) and (\ref{gamma-k'}))}\\
& = & (1-\epsilon')(\ln n)\gamma(G)\,, &
\end{array}$$
where $\epsilon' = \epsilon^3>0$.
Therefore, $S$ approximates domination within a factor of
$(1-\epsilon')\ln n$. By Theorem~\ref{thm:D-inapprox}, this shows that there is no polynomial time algorithm approximating
$k$-domination within a factor of \hbox{$(1-\epsilon)\ln n$}, unless
${\sf NP}\subseteq {\sf DTIME}(n^{O(\log\log n)})$.
\end{sloppypar}
\end{proof}

{A minor modification of the above proof (perform a reduction from total domination instead of domination)
can be used to obtain the analogue of Theorem~\ref{thm:k-domination-inapprox} for {$k$-tuple total domination}.

\begin{sloppypar}
\begin{thm}
\label{thm:k-tuple-total-domination-inapprox}
For every $k\ge 1$ and every $\epsilon>0$,  there is no polynomial time algorithm approximating
$k$-tuple total domination within a factor of \hbox{$(1-\epsilon)\ln n$}, unless
${\sf NP}\subseteq {\sf DTIME}(n^{O(\log\log n)})$.
\end{thm}
\end{sloppypar}

\subsection{Inapproximability under the ${\sf P} \neq {\sf NP}$ assumption}

Alon {\em et al.}~proved in~\cite{AMS06} that there is no polynomial time algorithm approximating
the set cover problem within a factor of \hbox{$c\ln n$} for some constant $c>0.2267$, unless ${\sf P} = {\sf NP}$.
Thus, one could obtain an analogue of Theorem~\ref{thm:D-inapprox} with a weaker inapproximability bound and under the assumption that
${\sf P} \neq {\sf NP}$.

In particular, with a straightforward adaptation of the proofs of the results in this section
(and, for $k$-tuple domination, of the proof by Klasing and Laforest~\cite{KL04}),
we obtain the following result.

\begin{sloppypar}
\begin{thm}\label{thm:P-NP-inapprox}
Unless ${\sf P} = {\sf NP}$, the following holds:\footnote{In the published version of this manuscript~\cite{CMV12}, lower bounds for vector domination and total vector domination of $0.1133\ln n$ and $0.0755\ln n$, respectively, were stated. However, since vector domination and total vector domination generalize domination and total domination, respectively, the better lower bound of $0.2267\ln n$ clearly holds for both problems.}
\begin{itemize}
  \item For every problem $\Pi\in \{$domination, $k$-domination, $k$-tuple domination,
  $k$-tuple total domination, multiple domination,
total domination, vector domination, total vector domination$\}$ (see Table~\ref{fig:table})
there is no polynomial time algorithm approximating $\Pi$ within a factor of \hbox{$0.2267\ln n$}.
  \item For every problem $\Pi\in \{\alpha$-domination,
  partial monopoly, strict $\alpha$-domination$\}$ (see Table~\ref{fig:table})
there is no polynomial time algorithm approximating $\Pi$ within a factor of \hbox{$0.1133\ln n$}.
  \item For every problem $\Pi\in \{\alpha$-rate domination,
monopoly, positive influence domination, strict total $\alpha$-domination,
total $\alpha$-domination$\}$ (see Table~\ref{fig:table})
there is no polynomial time algorithm approximating $\Pi$ within a factor of \hbox{$0.0755\ln n$}.
\end{itemize}
\end{thm}

The following corollary is immediate.

\begin{cor}\label{cor:NP}
All the problems listed in Theorem~\ref{thm:P-NP-inapprox} are {\sf NP}-complete.
\end{cor}

To the best of our knowledge, this is the first ${\sf NP}$-completeness proof for
$\alpha$-rate domination, $k$-tuple total domination, strict $\alpha$-domination, strict total $\alpha$-domination, and total $\alpha$-domination. For references to {\sf NP}-completeness proofs of the remaining problems, see Table~\ref{fig:table2}.
\end{sloppypar}

\section{Polynomial algorithms for particular graph classes}

In this section, we present several polynomial time algorithms for vector domination and total vector domination in particular graph classes.
For notational convenience, we will often replace in this section the vector notation ${\bf k} = (k_v:v\in V)$ with the function notation: an instance to the (total) vector domination problem will be given by a pair $(G,k)$ where $G = (V,E)$ is a graph and $k:V\longrightarrow \mathbb{N}_0$ is a function.

We start with complete graphs.

\subsection{Complete graphs}

\begin{sloppypar}
\begin{prop}
Let $G$ be a complete graph with vertex set $V(G) = \{v_1,\ldots, v_n\}$ and assume that
$n-1\ge k(v_1)\ge \cdots \ge k(v_n)\ge k(v_{n+1}):= 0$ with $k(v_1)>0$.
Then, a minimum vector dominating set for $(G,k)$ is given by $D = \{v_1,\ldots, v_p\}$
where $p = \min\{i\,:\,1\le i \le n,\, i\ge k(v_{i+1})\}$.
\end{prop}
\end{sloppypar}

\begin{proof}
Clearly, $D= \{v_1,\ldots, v_p\}$ as above  is a vector dominating set for $(G,k)$ since every $v\in V(G)\setminus D$ is of the form
$v = v_j$ for some $j\ge p+1$ and therefore $|N(v_j)\cap D| = |D| = p\ge k(v_{p+1})\ge k(v_j)$.
Conversely, if $D$ is a set of at most  $p-1$ vertices, then there exists a vertex $v_i\in V(G)\setminus D$ such that
$i\le p$. By definition of $p$, we have $p-1< k(v_{p})$.
Therefore, $|N(v_i)\cap D| \le p-1< k(v_p)\le k(v_i)$, hence $D$ is not a vector dominating set for $(G,k)$.
\end{proof}

\begin{cor}
For complete graphs, the vector domination problem is solvable in time $O(n)$.
\end{cor}

The claimed time bound can be achieved as follows: Since all the $k(v_i)$'s are less than $n$, they can be sorted using counting sort
in time $O(n)$. Thus, the value of $p = \min\{i\,:\,1\le i \le n,\, i\ge k(v_{i+1})\}$ and with it a minimum vector dominating set can also be computed within this time bound.

Total vector domination is even simpler and also solvable in $O(n)$ time.

\begin{sloppypar}
\begin{prop}\label{prop:complete-TVD}
Let $G=(V,E)$ be a complete graph. Let $K=\max\{k(v)\,:\,v\in V(G)\}$
and let $M = \{v\in V\,:\,k(v) = K\}$.
If $|M|\le |V|-K$, then a minimum total vector dominating set for $(G,k)$ is given by
any subset of $K$ vertices contained in $V\setminus M$.
Otherwise, a minimum total vector dominating set for $(G,k)$ is given by any subset of $K+1$ vertices.
\end{prop}
\end{sloppypar}

\begin{proof}
Clearly, every total vector dominating set must contain at least $K$ vertices.
Suppose first that $|M|\le |V|-K$, and let $D$ be a
subset of $K$ vertices contained in $V\setminus M$. Then, $k(v)<M$ for every $v\in D$, implying that
$|N(v)\cap D| = |D|-1 = M -1\ge k(v)$. For every $v\in V\setminus D$, we have
$|N(v)\cap D| = |D| = M \ge k(v)$.

Suppose now that $|M|>|V|-K$. In this case,
every subset $D\subseteq V$ with exactly $K$ vertices contains an element from $M$, say $v$, therefore
$|N(v)\cap D| = |D|-1 = K-1 < k(v)$. It follows that every total vector dominating set must contain at least $K+1$ vertices.
On the other hand, since $G$ is complete, every set $D$ with $K+1$ vertices will contain at least $K$ neighbors of every vertex.
\end{proof}

\begin{cor}
For complete graphs, the total vector domination problem is solvable in time $O(n)$.
\end{cor}

The claimed time bound can be achieved as follows: Assuming, as usual, that comparing two numbers can be done in constant time, computing the maximum $\max\{k(v)\,:\,v\in V(G)\}$ and the set $M$ can be done in time $O(n)$.
     The rest follows using Proposition~\ref{prop:complete-TVD}.

\subsection{Trees}\label{sec:trees}

Total vector domination and multiple domination problems are solvable in linear time on trees~\cite{LC-02} and also
in the larger class of strongly chordal graphs~\cite{LC-03,Uehara-02}. However, in \cite{LC-03}
Liao and Chang mention that their approach cannot be modified to handle the case of vector domination, and that a new approach
is needed. In this section we describe a linear time algorithm that solves vector domination in trees.


Given a tree $T$, we root it at an arbitrary vertex $r$. For a vertex $v$ of $T,$
we denote with $T_v$ the subtree of $T$ rooted at $v.$ For each vertex $v\neq r$ we also use $p(v)$ to denote the parent of $v,$ i.e.,
the last vertex ($v$ excluded) in the unique path from the root of $T$ to $v,$ and by $C(v)$ the set of children of $v$.

\begin{algorithm}[h!]
\caption{Vector domination in trees}\label{algo:trees}
Input: A tree $T=(V,E)$, a function $k:V\longrightarrow \mathbb{N}_0$.\\
Output: A set $S \subseteq V$ which is a minimum vector dominating set for~$(T,k)$.\\
\vspace{-0.4cm}
\begin{algorithmic}[1]
\STATE Fix a root $r\in V(T)$, and let $v_1,\ldots, v_n$ be the vertices of $T$ listed
in reverse order with respect to the time they are visited by  a breadth-first traversal from $r$.
\STATE Set $S = P = \emptyset$.
\FORALL{$i = 1, \ldots, n$}
\IF{$v_i\not\in P$}
\STATE $P\leftarrow P\cup \{v_i\}$
\IF{$i <n$ (i.e., $v_i$ is not the root)}
\IF{$|C(v_i)\cap S|\le k(v_i)-2$}\label{if1}
\STATE $S\leftarrow S\cup \{v_i\}$
\ELSIF{$|C(v_i)\cap S|=k(v_i)-1$}\label{if2}
\STATE $S\leftarrow S\cup \{p(v_i)\}$
\STATE $P\leftarrow P\cup \{p(v_i)\}$
\ENDIF
\ELSIF{$|C(v_i)\cap S|<k(v_i)$}\label{if3}
\STATE $S\leftarrow S\cup \{v_i\}$
\ENDIF
\ENDIF
\ENDFOR
\RETURN $S$
\end{algorithmic}
\end{algorithm}

\begin{thm}\label{thm:trees}
A minimum vector dominating set in a tree can be found in time $O(n)$.
\end{thm}

\begin{proof}
We claim that Algorithm~\ref{algo:trees} computes a minimum vector dominating set for~$(T,k)$.
The algorithm traverses the tree bottom up, processing vertices one at a time
in the opposite order as they are encountered by a breadth-first traversal from the root.
If, by the time a node $v_i\neq r$ is processed, at most
$k(v_i)-2$ children of $v_i$ have been put in $S$ (line~\ref{if1})
then there is no other way to satisfy the requirement for $S$ to be
a vector dominating set than to add $v_i$ to $S$,
and accordingly the algorithm does so.
If $k(v_i)-1$ children of $v_i$ have been put
in $S$ (line~\ref{if2}), then the additional neighbor of $v_i$
to include in $S$ is chosen as $v_i$'s parent $p(v_i)$,
for obvious reasons.
If $k(v_i)$ (or more) children of $v_i$ have already been put into $S$, then
there is nothing to do.
Line~\ref{if3} takes care of the limit case in which $v_i=r$ and therefore no
parent of $v_i$ exists.


For $i\in \{0,1,\ldots, n\}$, let us denote by $P_i$ and $S_i$ the sets $P$ and $S$, respectively,
after $i$ iterations of the {\bf for} loop. (In particular, $S_0 = P_0 = \emptyset$.)
To show the correctness of the algorithm, it suffices to prove the following claim:

{\it Claim:}
For every $i\in \{0,1,\ldots, n\}$, the following holds:

$(a)$ For every $v\in P_i\setminus S_i$, it holds that $|N(v)\cap S_i|\ge k(v)$.

$(b)$ There exists a minimum vector dominating set $D_i$ for $(T,k)$ such that $D_i\cap P_i = S_i$.

\medskip
The correctness of the algorithm follows from part $(b)$ of the claim for $i = n$.

We prove the claim by induction on $i$. Both statements hold trivially  for $i = 0$.

For the inductive step, we consider the two statements separately. First we consider part $(a)$.

Let $i\ge 1$ and suppose that the statement in $(a)$ holds for smaller values of~$i$.
If $P_{i}\setminus S_{i} = P_{i-1}\setminus S_{i-1}$, the statement holds by the induction hypothesis,
since $S_{i-1}\subseteq S_{i}$. There are only two remaining cases:

(i) $S_{i} = S_{i-1}\cup \{p(v_{i})\}$ and $P_{i} = P_{i-1}\cup \{v_{i}, p(v_{i})\}$.
This corresponds to the case when the {\bf if} statement in line~\ref{if2} is true.
In this case $P_{i}\setminus S_{i} = (P_{i-1}\setminus S_{i-1})\cup \{v_{i}\}$, and we only need to verify
the condition $|N(v_{i})\cap S_{i}|\ge k(v_{i})$. This is indeed the case since by the standing hypothesis,
exactly $k(v_{i})-1$ of the children of $v_{i}$ are in $S_{i-1}$, and $S_{i}$ contains all of them plus the parent of $v_{i}$.

(ii) $S_{i} = S_{i-1}$ and $P_{i} = P_{i-1}\cup \{v_{i}\}$.
In this case $P_{i}\setminus S_{i} = (P_{i-1}\setminus S_{i-1})\cup \{v_{i}\}$, and we only need to verify
the condition
$|N(v_{i})\cap S_{i}|\ge k(v_{i})$.
This case only happens when none of the conditions in lines~\ref{if1}, \ref{if2} and \ref{if3} are satisfied, which means that
$S_{i-1}$ contains at least $k(v_{i})$ neighbors, in fact, children, of $v_{i}$. Thus, the condition holds
by the induction hypothesis.

Now we consider part $(b)$. Let $i\ge 1$ and suppose that the statement in $(b)$ holds for smaller values of~$i$.

We may assume that $P_{i} \neq P_{i-1}$ since otherwise we also have $S_{i} = S_{i-1}$ and we can take
$D_i = D_{i-1}$, applying the inductive hypothesis.

Suppose first that $S_{i} = S_{i-1}$. Then none of the conditions in lines~\ref{if1}, \ref{if2} and \ref{if3} are satisfied, which means that
$S_{i-1}$ contains at least $k(v_{i})$ children of $v_{i}$.

If $v_i\not\in D_{i-1}$, then we can take $D_i = D_{i-1}$.

If $v_i\in D_{i-1}$, then by the minimality of $D_{i-1}$, we see that $v_{i}\neq r$ and
$p(v_i)\not\in D_{i-1}$, for otherwise we could remove $v_i$ from $D_{i-1}$ to obtain a smaller vector dominating set for $(T,k)$.
Let $D_i = (D_{i-1}\setminus \{v_i\})\cup \{p(v_i)\}$.
Clearly, $|D_i| = |D_{i-1}|$ and $D_i\cap P_i = S_i$. Now we argue that $D_i$ is a vector dominating set for $(T,k)$.
Notice that $D_{i-1}\setminus\{v_i\}$ vector dominates all vertices of $T$ except possibly $p(v_i)$:
for $v_i$, this is the case due to the fact that $S_{i-1}$ is included in $D_{i-1}\setminus \{v_i\}$ (by the induction hypothesis),
and $S_{i-1}$ contains at least $k(v_{i})$ children of $v_{i}$; moreover, $S_{i-1}$ vector dominates
all children of $v_i$ by $(a)$; finally, for every vertex outside $T_{v_i}\cup\{p(v_i)\}$ this is true because it is
vector dominated by $D_{i-1}$ and it is not a neighbor of $v_i$.

Now suppose that $S_{i} \neq S_{i-1}$.

If $S_i = S_{i-1}\cup \{p(v_i)\}$ then  the condition in line \ref{if2} is true.
Since $C(v_i) \subseteq P_{i-1}$ we have that
  $C(v_i) \cap D_{i-1} =  C(v_i) \cap (P_{i-1} \cap D_{i-1}) =  C(v_i) \cap S_{i-1}$
  where the last equality follows by the inductive hypothesis. Thus
$|C(v_i)\cap D_{i-1}| =  |C(v_i)\cap S_{i-1}| = k(v_{i})-1.$
Hence, if $D_{i-1}$ does not  contain $v_i$ then it must contain  $p(v_i).$
If  $v_i  \not \in D_{i-1}$ then  it is easy to see that taking $D_i = D_{i-1}$ satisfies the claim.  On the other hand, if
$v_i  \in D_{i-1}$ then we  can set $D_i = (D_{i-1}\setminus \{v_i\})\cup \{p(v_i)\}.$
Clearly, $|D_i| \leq |D_{i-1}|$ and $D_i\cap P_i = S_i$.
Notice that $D_{i-1}\setminus\{v_{i}\}$---hence also $D_i$---vector dominates all vertices of $T_{v_{i}} - v_i$ by $(a)$ and the induction hypothesis.
Also, because of  $|D_{i-1} \cap C(v_i)| = k(v_{i})-1,$ we have that $D_i $ vector dominates $v_i.$ Finally, $p(v_i)$ is trivially vector dominated by $D_i$, and
so is every remaining  vertex $v$ of $T$ because $D_i \cap N(v) \supseteq D_{i-1} \cap N(v).$

Finally, if $S_i = S_{i-1}\cup \{v_i\}$ then we can take $D_i = D_{i-1}$. It suffices to argue that $D_{i-1}$ contains $v_i$.
Again, since $C(v_i) \subseteq P_{i-1}$ we have that
  $C(v_i) \cap D_{i-1} =  C(v_i) \cap (P_{i-1} \cap D_{i-1}) =  C(v_i) \cap S_{i-1}$
  where the last equality follows by the inductive hypothesis.
By the standing assumption the set $C(v_i)\cap D_{i-1}$
is too small to allow $D_{i-1}$ to vector dominate $v_i,$ unless $v_i\in D_{i-1}$.

This completes the proof of the claim and with it the proof of the correctness of the algorithm.

\medskip

It remains to analyze the time complexity of the algorithm. A breadth-first traversal together with computing the parents $p(v)$ can be done in linear time. It is not hard to see that all the operations performed by the algorithm at any vertex $v_i$ take constant time:
the only operation that requires some care is the computation of
the cardinality  of set intersection $C(v_i)\cap S$ needed in lines
$\ref{if1}, \ref{if2}, \ref{if3}$. For this, we keep a counter for each vertex, which is originally set to $0$;
morevoer, every time we include a new vertex into $S$, we increase by $1$ the counter of its parent (if it exists).
Therefore, the algorithm can be implemented to run in linear time.

\end{proof}

\subsection{$P_4$-free graphs}

\begin{sloppypar}
In this section we give a polynomial time algorithm to solve the vector domination and total vector domination problems in $P_4$-free graphs.
$P_4$-free graphs (also known as cographs) are graphs without an induced subgraph isomorphic to a $4$-vertex path.
A polynomial time algorithm for $k$-tuple domination in a class of graphs properly containing the $P_4$-free graphs was recently given in~\cite{DLN-11}.

 The algorithm will be based on the following well-known characterization of $P_4$-free graphs~\cite{CLB81}: a graph $G$ is $P_4$-free if and only if for every induced subgraph $F$ of $G$ with at least two vertices, either $F$ or its complement is disconnected. A \emph{co-component} of a graph $G=(V,E)$ is the subgraph of $G$ induced by the vertex set of a connected component of the complementary graph \hbox{$\overline G = (V,\{uv~|~u,v\in V,~u\neq v,~uv\not\in E\})$}.
The above characterization implies that every $P_4$-free graph $G=(V,E)$ admits a recursive decomposition into one-vertex graphs by taking components or co-components. Such a decomposition can be computed in linear time~\cite{CPS85}, and a tree representing such a decomposition is called a {\em cotree}. For our purposes, it will be more convenient to assume that $G$ is represented by a {\em modified cotree}, which is obtained from the cotree by replacing every
node representing a decomposition of an induced subgraph $F$ of $G$ into $p\ge 3$ co-components $C_1,\ldots, C_p$
with $p-1$ nodes in sequence, with $i$-th node representing the decomposition of $F_i := F-(C_1\cup\cdots\cup C_{i-1})$ into $C_i$ and $F_i-C_i$.
\end{sloppypar}

\begin{prop}\label{prop:join}
Let $G$, $G_1$, $G_2$ be graphs such that $G$ is obtained from the disjoint union of $G_1$ and $G_2$ by adding all edges of the form
$\{uv~:u\in V(G_1),~v\in V(G_2)\}$. Then, 
$$\gamma(G,k)= \min_{\substack{0\le i\le |V(G_2)|\\0\le j\le |V(G_1)|\\}}\left(\max\{\gamma(G_1,k_i), j\} + \max\{\gamma(G_2,k_j'), i\}\right)$$

$$\gamma^t(G,k)= \min_{\substack{0\le i\le |V(G_2)|\\0\le j\le |V(G_1)|\\}}\left(\max\{\gamma^t(G_1,k_i), j\} + \max\{\gamma^t(G_2,k_j'), i\}\right)\,,$$
where $k_i(v) = \max\{k(v)-i,0\}$ for all $v\in V(G_1)$ and $k_j'(v) = \max\{k(v)-j,0\}$ for all $v\in V(G_2)$.
\end{prop}

\begin{proof}
\begin{sloppypar}
Let $m$ denote the value of the first minimum above. First, we show that $m\le \gamma(G,k)$. Let $D$ be a minimum vector dominating set for $(G,k)$, that is, $|D| = \gamma(G,k)$.
Let $D_i = D\cap V(G_i)$, for $i = 1,2$, and let $i^* = |D_2|$ and $j^* = |D_1|$.
Take a vertex $v\in V(G_1)\setminus D_1$ such that $k_{i^*}(v)>0$. Then
$$|N_{G_1}(v)\cap D_1| =|N_{G}(v)\cap D| -|D_2| = |N_{G}(v)\cap D| - i^* \ge k(v)-i^* = k_{i^*}(v)\,.$$ Therefore
$D_1$ is a vector dominating set for $(G_1,k_{i^*})$ and consequently  $\gamma(G_1,k_{i^*})\le |D_1| = j^*$.
Similarly, we can show that $\gamma(G_2,k_{j^*}')\le |D_2| = i^*$.
It follows that $$\gamma(G,k) = |D| = j^*+i^*= \max\{\gamma(G_1,k_{i^*}), j^*\} + \max\{\gamma(G_2,k_{j^*}'), i^*\}\ge m\,.$$
To see the converse inequality, let $(i^*, j^*)$ be a pair of indices where the value of  $m$ is attained.
Let $D_1$ be a vector dominating set for  $(G_1, k_{i^*})$ such that $|D_1| = \max\{\gamma(G_1,k_{i^*}), j^*\}$.
Similarly, let $D_2$ be a vector dominating set for $(G_2, k_{j^*}')$ such that $|D_2| =
\max\{\gamma(G_2,k_{j^*}), i^*\}$. Then, the set $D := D_1\cup D_2$ is a vector dominating set for $(G,k)$:
Let $v\in V(G)\setminus D$. Assuming that $v\in V(G_1)\setminus D_1$, we have
$$|N_G(v)\cap D| = |N_{G_1}(v)\cap D_1| + |D_2| \ge k_{i^*}(v) + |D_2| \ge k(v)-i^*+|D_2|\ge k(v)\,.$$
We can show similarly that $|N_G(v)\cap D| \ge k(v)$ for all $v\in V(G_2)\setminus D_2$. Therefore, $\gamma(G,k)\le |D| = |D_1|+|D_2| = m$, which completes the proof.

The proof of the other relation is analogous.
\end{sloppypar}
\end{proof}

\begin{thm}\label{thm:P_4-free}
Vector domination problem and total vector domination problem are solvable in polynomial time on $P_4$-free graphs.
\end{thm}

\begin{proof}
We claim that Algorithm~\ref{algo:P4-free} below computes a minimum vector dominating set for $(G,k)$, where $G$ is a
$P_4$-free graph. The following notations are used:
For a non-negative integer $r$ and for an induced subgraph $H$ of $G$,
we denote by $D(H,r)$ a minimum vector dominating set
for $(H,k_r)$, where $k_r(v) = \max\{k(v)-r,0\}$ for all $v\in V(H)$. 

\begin{algorithm}[h!]
\caption{Vector domination in $P_4$-free graphs}\label{algo:P4-free}
Input: A $P_4$-free graph $G=(V,E)$, a function $k:V\longrightarrow \mathbb{N}_0$.\\
Output: A minimum vector dominating set for $(G,k)$.\\
\vspace{-0.4cm}
\begin{algorithmic}[1]
\STATE Let $R = \{v\in V(G)~:k(v)> d(v)\}$.
\STATE Set $G$ to $G-R$ and $k$ to $k':V(G-R)\longrightarrow \mathbb{N}_0$, given by $k'(v) = \max\{k(v) - |N(v)\cap R|,0\}$ for all $v\in V(G)-R$.
\STATE Compute a modified cotree $T$ of $G$.
\FORALL{leaves $\ell$ of $T$}
\STATE let $v\in V(G)$ be the vertex corresponding to $\ell$.
\FORALL{$0\le r \le \Delta(G)$}
\STATE set $D(\{v\},r)= \left\{
                                               \begin{array}{ll}
                                                 \emptyset, & \hbox{if $k(v) \le r$;} \\
                                                 \{v\}, & \hbox{otherwise.}
                                               \end{array}
                                             \right.$
\ENDFOR
\ENDFOR
\FORALL{internal nodes of $T$ (traversed in a bottom-up manner)}
\STATE let $H$ be the subgraph of $G$ corresponding to the current node of $T$.
\IF{$H$ is disconnected, with connected components $C_1,\ldots, C_m$}
\FORALL{$0\le r \le \Delta(G)$}
\STATE set $D(H,r)=\cup_{1\le i\le m} D(C_i, r)\,.$
\ENDFOR
\ELSE
\STATE let $C$ be a co-component of $H$ and let $H_2 = H-C$.
\FORALL{$0\le r \le \Delta(G)$}
\FORALL{$0\le i \le |V(H_2)|$}
\STATE let $D_i = D(C,\min\{r+i,\Delta(G)\})$.
\ENDFOR
\FORALL{$0\le j \le |V(C)|$}
\STATE let $D_j' = D(H_2,\min\{r+j,\Delta(G)\})$.
\ENDFOR
\STATE let $(i^*,j^*)$ be a pair of indices such that
$\max\{|D_{i^*}|,j^*\}+\max\{|D_{j^*}'|,i^*\}=\min_{i,j} \left(\max\{|D_{i}|,j\}+\max\{|D_{j}'|,i\}\right)$
\STATE let $\hat D_1 = D_{i^*}\cup J$ where $J\subseteq V(C)\setminus D_{i^*}$ such that $|J| = \max\{j^*-|D_{i^*}|,0\}$.
\STATE let $\hat D_2 = D_{j^*}'\cup J$ where $J\subseteq V(G_2)\setminus D_{j^*}'$ such that $|J| = \max\{i^*-|D_{j^*}'|,0\}$.
\STATE set $D(H,r) =\hat D_1\cup \hat D_2$.
\ENDFOR
\ENDIF
\ENDFOR
\RETURN $D(G,0)\cup R$.
\end{algorithmic}
\end{algorithm}

In lines 1--2, the algorithm computes the set $R$ of required vertices in every feasible solution, and reduces the problem to a smaller graph.
Notice that once the required vertices have been removed, it holds that $k(v)\le d(v)$ for all $v$. In particular, for an induced subgraph $H$ of the reduced graph $G-R$, it suffices to compute the sets $D(H,r)$ for $r\le \Delta(G)$, since
$D(H,r') = \emptyset$ for every $r'\ge \Delta(G)$.

\begin{sloppypar}
The correctness of the algorithm is straightforward, using the above-mentioned characterization of $P_4$-free graphs and Proposition~\ref{prop:join} together with the arguments given in its proof. It is also easy to see that the
algorithm runs in  time $O(\Delta(G)n^3)$.
\end{sloppypar}

The algorithm can be modified slightly so that it computes a minimum total vector dominating set. Suppose that
an induced subgraph $H$ of $G$ contains a vertex $v$ such that $k(v)-r> d(v)$. In this case, we set $D(H,r) = \textrm{Inf}$ where $\textrm{Inf}$ is a special symbol denoting
the infeasibility of the problem (we also set $|\textrm{Inf}| = \infty$); moreover $\textrm{Inf}$ is invariant under taking unions: $A\cup \textrm{Inf} = \textrm{Inf}$ for every $A$.
We need the following modifications:
\begin{itemize}
\item replace lines 1--2 with the following:

{\bf if}~~there exists a vertex $v$ such that $k(v)>d(v)$ {\bf then}~~return \textrm{Inf}.
\item replace line 7 with the following:

set $D(\{v\},r)= \left\{
                                               \begin{array}{ll}
                                                 \emptyset, & \hbox{if $k(v) \le r$;} \\
                                                 \textrm{Inf}, & \hbox{otherwise.}
                                               \end{array}
                                             \right.$
\end{itemize}

\end{proof}


\subsection{Threshold graphs}

Threshold graphs form a subclass of $P_4$-free graphs, therefore vector domination and total vector domination problems are solvable
in polynomial time on threshold graphs. Since threshold graphs are strongly chordal, the total vector domination problem
is solvable in time $O(n+m)$ on threshold graphs~\cite{LC-03,Uehara-02}. We develop in this section an $O(nm)$ algorithm for the vector domination problem in threshold graphs, using the following characterization:
A graph $G = (V,E)$ is threshold if and only if there is an ordering $v_1,\ldots, v_n$ of $V$ such that for every $i$, vertex $v_i$
is either isolated or dominating in the subgraph $G_i$ of $G$ induced by $\{v_1,\ldots, v_i\}$. Such an ordering of
a threshold graph $G$ can be found in linear time by recursively removing dominating or isolated vertices.


We will also need the following proposition similar to Proposition~\ref{prop:join}.
For a subgraph  $H$ of $G$, we denote by $k|_{H}$ the restriction of $k$ to $V(H)$, that is, the function $k|_{H}:V(H)\longrightarrow\mathbb{N}_0$, given by
$k|_{H}(v) = k(v)$ for all $v\in V(H)$.

\begin{prop}\label{prop:dominating}
Let $G$ be a graph with a dominating vertex $v$.
Let $G' = G-\{v\}$ and $k':V(G')\longrightarrow\mathbb{N}_0$ be given by
$k'(u) = \max\{k(u)-1,0\}$ for all $u\in V(G')$.
If $k(v)>d(v)$ then every
minimum vector dominating set $D$ for $(G,k)$ is of the form $D'\cup \{v\}$ where $D'$ is a
minimum vector dominating set for $(G',k')$.
Otherwise,
$$\gamma(G,k) = \min\{\max\{\gamma(G',k|_{G'}),k(v)\},1+\gamma(G',k')\}\,.$$
More specifically,
if $D'$ is a minimum vector dominating set for $(G',k|_{G'})$
and $D''$ is a minimum vector dominating set for $(G',k')$ then
a minimum vector dominating set $D$ for $(G,k)$ can be computed as follows:
$$D = \left\{
       \begin{array}{ll}
         D'\cup J, & \hbox{if $\max\{|D'|,k(v)\}\le 1+\gamma(G',k')$;} \\
         D''\cup \{v\}, & \hbox{otherwise,}
       \end{array}
     \right.$$
where $J\subseteq V(G')\setminus D'$ such that $|J| = \max\{k(v) -|D'|,0\}$.
\end{prop}

\begin{proof}
If $k(v)>d(v)$ then every   minimum vector dominating set $D$ for $(G,k)$ must contain $v$, and the first statement follows.

Suppose now that $k(v)\le d(v)$. Let $D$ be a minimum vector dominating set for $(G,k)$.
If $v\in D$ then $D' = D\setminus \{v\}$ is a vector dominating set for $(G',k')$. Therefore, in this case
$\gamma(G',k')\le \gamma(G,k)-1$ and the inequality
$\gamma(G,k) \ge \min\{\max\{\gamma(G',k|_{G'}),k(v)\},1+\gamma(G',k')\}$
follows.
If $v\not\in D$ then $D' = D\setminus \{v\}$ is a vector dominating set for $(G',k|_{G'})$, moreover
$|D'|\ge k(v)$; therefore the inequality
$\gamma(G,k) \ge \min\{\max\{\gamma(G',k|_{G'}),k(v)\},1+\gamma(G',k')\}$
holds in this case too.

\begin{sloppypar}
To see the converse inequality, suppose first that $\max\{\gamma(G',k|_{G'}),k(v)\}\le 1+\gamma(G',k')$, and let
$D'$ be a minimum vector dominating set for $(G',k|_{G'})$. Let $D = D'\cup J$ where $J\subseteq V(G')\setminus D'$ such that $|J| = \max\{k(v) -|D'|,0\}$.
Then, the set $D$ contains at least $k(v)$ neighbors of $v$, therefore $D$ is a
vector dominating set for $(G,k)$. Similarly, if $\max\{\gamma(G',k|_{G'}),k(v)\}> 1+\gamma(G',k')$, then letting $D''$
be a minimum vector dominating set for $(G',k')$, we can define $D = D''\cup \{v\}$ to obtain a vector dominating set for $(G',k')$.
In summary, $\gamma(G,k) \le \min\{\max\{\gamma(G',k|_{G'}),k(v)\},1+\gamma(G',k')\}$; hence equality holds, and the set $D$ is also a minimum
vector dominating set for $(G,k)$.
\end{sloppypar}
\end{proof}

Proposition~\ref{prop:dominating} leads to Algorithm~\ref{algo:threshold} below for the vector domination problem on threshold graphs.

\begin{algorithm}[h!]
\caption{Vector domination in threshold graphs}\label{algo:threshold}
Input: A threshold graph $G=(V,E)$, a function $k:V\longrightarrow \mathbb{N}_0$.\\
Output: A minimum vector dominating set for $(G,k)$.\\
\vspace{-0.4cm}
\begin{algorithmic}[1]
\STATE Let $R = \{v\in V(G)~:k(v)> d(v)\}$.
\STATE Set $G$ to $G-R$ and $k$ to $k':V(G-R)\longrightarrow \mathbb{N}_0$, given by $k'(v) = \max\{k(v) - |N(v)\cap R|,0\}$ for all $v\in V(G)-R$.
\STATE Compute an ordering $v_1,\ldots, v_n$ of $V(G)$ such that $v_i$ is either isolated or dominating in $G_i$.
\STATE Compute the values $p_j$ for all $j\in \{1,\ldots, n\}$.
\FORALL {$0\le j\le p_1$}
\STATE set $D_{1,j} = \left\{
                         \begin{array}{ll}
                           \emptyset, & \hbox{if $k(v_1)\le j$;} \\
                           \{v_1\}, & \hbox{otherwise.}
                         \end{array}
                       \right.$
\ENDFOR
\FORALL{$i = 2,\ldots, n$}
\IF{$v_i$ is isolated in $G_i$}
\FORALL{$0\le j\le p_i$}
\STATE set $D_{i,j}=\left\{
                                     \begin{array}{ll}
                                       D_{i-1,j}, & \hbox{if $k(v_i)\le j$;} \\
                                       D_{i-1,j}\cup\{v_i\}, & \hbox{otherwise.}
                                     \end{array}
                                   \right.$
\ENDFOR
\ELSE
\FORALL{$0\le j\le p_i$}
\IF{$\max\{|D_{i-1,j}|, k(v)-j\} \le 1 + |D_{i-1,j+1}|$}
\STATE let $J\subseteq V(G_{i-1})\setminus D_{i-1,j}$ such that $|J| = \max\{k(v)-j - |D_{i-1,j}|, 0\}$.
\STATE set $D_{i,j} = D_{i-1,j}\cup J$.
\ELSE
\STATE set $D_{i,j} = D_{i-1,j+1}\cup \{v_i\}$.
\ENDIF
\ENDFOR
\ENDIF
\ENDFOR
\RETURN $D_{n,0}\cup R$.
\end{algorithmic}
\end{algorithm}

\begin{thm}\label{thm:threshold}
A minimum vector dominating set in a threshold graph can be found in time $O(nm)$.
\end{thm}

\begin{proof}
We claim that Algorithm~\ref{algo:threshold}  
 computes a minimum vector dominating set for $(G,k)$, where $G$ is a
threshold graph. We use similar notation as in the proof of Theorem~\ref{thm:P_4-free}, except that
we denote by $D_{i,j}$ a minimum vector dominating set for $(G_i, k_j)$ where
$k_j(v) = \max\{k(v)-j,0\}$ for all $v\in V(G_i)$. The algorithm will compute, by dynamic programming,
all sets $D_{i,j}$, for all $i \in \{1,\ldots, n\}$ and all $j\in \{0,1,\ldots, p_i\}$
where $p_i$ is the number of indices $j>i$ such that $v_j$ is dominating in $G_j$.

The correctness of the algorithm follows by induction on $i$, using Proposition~\ref{prop:dominating}. Notice that for all $i\ge 2$ such that $v_i$
is dominating in $G_i$, we have $p_{i-1}= p_i+1$, therefore $j+1\le p_{i-1}$ in lines 15 and 19, so $D_{i-1,j+1}$ has already
been computed at that point. The total time complexity is $O(n\sum_{i = 1}^np_i) = O(nm)$, and can be improved to $O(n+m)$ if only the
minimum size of a vector dominating set is needed.
\end{proof}
%

\remove{
\section{Graphs of bounded clique-width}
}

\section{Concluding remarks}

We have studied some algorithmic issues related to
natural extensions of the well known concepts of domination
and total domination in graphs. We have shown that the problems
are approximable to within a logarithmic factor, and
proved that this is essentially best possible.
We summarize our and other known results in Table~\ref{fig:table2} (which should be read in conjuction with Table~\ref{fig:table}).
In the last column, we provide a reference for {\sf NP}-completeness proofs of the corresponding decision problems.\footnote{In the published version of this manuscript~\cite{CMV12}, lower bounds for vector domination and total vector domination of $(1/2-\epsilon)\ln n$ and $(1/3-\epsilon)\ln n$, respectively, were stated. However, since vector domination and total vector domination generalize domination and total domination, respectively, the better lower bound of $(1-\epsilon)\ln n$ clearly holds for both problems.}
\begin{center}
\begin{table}[h!]
  \centering
  \small
    \begin{tabular}{|l|c|c|c|}
  \hline
  { Model} & { Upper bound} & { Lower bound} & { {\sf NP}-completeness}\\
  \hline
  $\alpha$-domination & $\ln(2\Delta(G))+1$& $(1/2-\epsilon)\ln n$ & \cite{DHLM-2000}\\
  \hline
  $\alpha$-rate domination & $\ln(\Delta(G))+1$ & $(1/3-\epsilon)\ln n$ & Corollary~\ref{cor:NP}\\
  \hline
  domination & $\ln(\Delta(G)+1)+1/2$~\cite{CC04,DF-1997}& $(1-\epsilon)\ln n$~\cite{CC04}& \cite{GJ79}\\
  \hline
  $k$-domination & $\ln(2\Delta(G))+1$& $(1-\epsilon)\ln n$& \cite{JP-89}\\
  \hline
  $k$-tuple domination & $\ln(\Delta(G)+1)+1$~\cite{KL04} & $(1-\epsilon)\ln n$~\cite{KL04}& \cite{LC-03}\\
  \hline
  $k$-tuple total domination  & $\ln(\Delta(G))+1$ & $(1-\epsilon)\ln n$& Corollary~\ref{cor:NP}\\
  \hline
  monopoly &  $\ln(\Delta(G)+1)+1$~\cite{Peleg-02} & $(1/3-\epsilon)\ln n$~\cite{MRS-2002} & \cite{MRS-2002,Peleg-02}\\
  \hline
  multiple domination &  $\ln(\Delta(G)+1)+1$ & $(1-\epsilon)\ln n$ & {\scriptsize $\begin{array}{c}
                                                                         \textrm{generalizes} \\
                                                                         \textrm{domination}
                                                                       \end{array}$}\\
  \hline
  partial monopoly  &  $\ln(2\Delta(G))+1$~\cite{Peleg-02}& $(1/2-\epsilon)\ln n$~\cite{MRS-2002} & \cite{MRS-2002,Peleg-02}\\
 \hline
  positive influence domination &  $\ln(\Delta(G))+1$~\cite{Wang11} &  $(1/3-\epsilon)\ln n$ & \cite{Wang11}\\
 \hline
  strict $\alpha$-domination  &  $\ln(2\Delta(G))+1$ & $(1/2-\epsilon)\ln n$ & Corollary~\ref{cor:NP}\\
 \hline
  strict total $\alpha$-domination &  $\ln(\Delta(G))+1$ & $(1/3-\epsilon)\ln n$ & Corollary~\ref{cor:NP}\\
   \hline
 total domination &  $\ln(\Delta(G))+1/2$~\cite{CC04,DF-1997} & $(1-\epsilon)\ln n$~\cite{CC04}  & \cite{HHS1-98}\\
\hline
  total $\alpha$-domination &  $\ln(\Delta(G))+1$ & $(1/3-\epsilon)\ln n$  & Corollary~\ref{cor:NP}\\
\hline
  total vector domination &  $\ln(\Delta(G))+1$ & $(1-\epsilon)\ln n$  & {\scriptsize $\begin{array}{c}
                                                                         \textrm{generalizes} \\
                                                                         \textrm{total domination}
                                                                       \end{array}$}\\
  \hline
  vector domination &  $\ln(2\Delta(G))+1$& $(1-\epsilon)\ln n$ & {\scriptsize $\begin{array}{c}
                                                                         \textrm{generalizes} \\
                                                                         \textrm{domination}
                                                                       \end{array}$}  \\
\hline
\end{tabular}
\caption{Known approximability results for different domination problems. Lower bounds hold unless
 ${\sf NP}\subseteq {\sf DTIME}(n^{O(\log\log n)})$. 
They also hold unless ${\sf P} = {\sf NP}$ but the constant must be multiplied by $0.2267$.
Unless stated otherwise, all the upper and lower bounds in the table are from this paper.}
\label{fig:table2}
\end{table}
\end{center}

\begin{sloppypar}
We have also  provided exact polynomial time algorithms
for several interesting classes of graphs, namely, complete
graphs, trees, $P_4$-free graphs and threshold graphs.
We leave it as a question for future research to determine the complexity status of the vector domination and
related problems for graphs of bounded tree-width or bounded clique-width.
\end{sloppypar}

\subsubsection*{Note added in proof}

Some of the results presented in this paper were obtained independently of us by Dinh, Shen, Nguyen and Thai~\cite{DSNT}.

\noindent
{\bf Acknowledgements}

The authors are grateful to Dieter Rautenbach for telling them about the notion of $\alpha$-domination.
The linear time algorithm for vector domination in trees given in Section~\ref{sec:trees}
was inspired by discussions with Andr\'{e} Nichterlein. The authors are also grateful to two anonymous referees
whose comments helped to improve the presentation of the paper.

\begin{sloppypar}
The first author was partially supported by a DAAD  grant, ref.~code  A/11/15927. The work of the second author was partially done during several visits at the Department of Informatics at the University of Salerno; the kind hospitality of the first and the third authors is greatly appreciated.
The second author was supported in part by ``Agencija za raziskovalno dejavnost Republike Slovenije", research program P1-0285 and research projects J$1$--$4010$, J$1$--$4021$ and N$1$--$0011$. The work of the third author was partially done while visiting the Mascotte team of INRIA at Sophia Antipolis. He wants to thank J.-C.~Bermond and D.~Coudert for their kind hospitality.
\end{sloppypar}

\newpage

\appendix

\section{Proof of Theorem~\ref{thm:alpha-rate-inapprox}}

\begin{thm-alpha}
For every $\alpha\in (0,1)$ and every $\epsilon>0$,
there is no polynomial time algorithm approximating
$\alpha$-rate domination within a factor of \hbox{$(\frac{1}{3}-\epsilon)\ln n$}, unless
${\sf NP}\subseteq {\sf DTIME}(n^{O(\log\log n)})$.
\end{thm-alpha}

\begin{proof}
Let $0<\alpha<1$ and $\epsilon\in (0,\frac{1}{3}).$
Let $B = \lceil\frac{\alpha}{1-\alpha}\rceil$.
We make a reduction from domination on graphs $G$ with $n$ vertices, none of which are isolated, such that
\begin{equation}\label{size-of-G-1'}
n \ge \max\left\{\sqrt{\frac{1-\alpha}{\alpha}},(B+1)^{2/\epsilon}\right\}\,,
\end{equation}
\begin{equation}\label{size-of-G-2'}
\lceil n^{1+\epsilon/3}\rceil\le n^{1+2\epsilon/3}\,,
\end{equation}
\begin{equation}\label{size-of-G-3'}
n+B\le n^{1+\epsilon/3}\,.\end{equation}
and
\begin{equation}\label{gamma'}
\gamma(G)\ge \frac{B}{\epsilon}\,.
\end{equation}
Clearly, these assumptions are without loss of generality since the inequalities in~(\ref{size-of-G-1'})--(\ref{size-of-G-3'}) are satisfied for all large enough $n$, while if the inequality (\ref{gamma'}) is violated, we can find an optimal solution in polynomial time by verifying all subsets of $V(G)$ of size less than $\frac{B}{\epsilon}$.

Let $G$ be a graph satisfying $(\ref{size-of-G-1'})$--$(\ref{gamma'})$.
Let $n = |V(G)|$ and $m:=\lceil n^{1+\epsilon/3}\rceil$.
We transform $G$ into a graph $G'$ as follows:
$G'$ consists of $mn$ disjoint copies of $G$, say $G_1,\ldots, G_{mn}$,
together with a complete graph $K$ on $Bmn$ vertices such that
$K$ is disjoint from the $mn$ copies of $G$.
(See Fig.~\ref{fig:transformation2}.)
To describe the remaining edges,
we first partition the vertex set of $K$ into $m$ equally-sized parts $K_1,\ldots, K_{m}$.
(In particular, $|K_i| = Bn$ for all $i=1,\ldots, n$.) Finally, for every $j\in\{1,\ldots, mn\}$,
we make every vertex $v\in V(G_j)$ adjacent to precisely $k_v$ vertices in $K_{\lceil j/n\rceil}$
where $k_v$ is an integer satisfying
$$\frac{k_v}{d_G(v)+k_v+1}<\alpha\le\frac{k_v+1}{d_G(v)+k_v+1}\,.$$
We can take $k_v = \lceil\frac{\alpha d_G(v)+\alpha-1}{1-\alpha}\rceil$.
Notice that since $0\le k_v\le Bn$, it is indeed possible to assign to every
$v\in V(G_j)$ precisely $k_v$ neighbors in $K_{\lceil j/n\rceil}$.
(This assignment is done in an arbitrary way.)

{\em Claim:} $mn\gamma(G)\le \gamma_{\times\alpha}(G') \le mn\gamma(G) + Bmn\,.$

\bigskip
Proof of Claim:

Let $S'$ be an optimal $\alpha$-rate dominating set in $G'$, that is, $|S'| = \gamma_{\times\alpha}(G')$.
For every $j= 1,\ldots, mn$, let $S'_j = S'\cap V(G_j)$
denote the part of $S'$ that belongs to to the $j$-th copy of $G$ in $G'$.
Pick an index $j^*\in \{1,\ldots, mn\}$ for which the size of $S_j'$ is smallest.
We argue that the set $S:=S_{j^*}$ is a dominating set
in $G_{j^*}$ (and thus in $G$). Indeed, suppose for contradiction that
there exists a vertex $v$ in $G_{j^*}$ such that $S$ misses the closed neighborhood of $v$.
Then $|N_{G'}[v]\cap S'|\le k_v$ while the size of the closed neighborhood of $v$ in $G'$ is equal to
$|N_{G'}[v]|= d_G(v)+1+k_v$.
Therefore $$\frac{|N_{G'}[v]\cap S'|}{|N_{G'}[v]|}\le \frac{k_v}{d_G(v)+1+k_v}<\alpha\,,$$
contrary to the assumption that $S'$ is  $\alpha$-rate dominating.
This implies that $\gamma(G)\le |S|$ and consequently
$mn\gamma(G)\le mn|S|\le \sum_{j = 1}^{mn}|S'_j|\le |S'| =\gamma_{\times \alpha}(G')$.


Conversely, let $S$ be an optimal dominating set in $G$. For $j = 1,\ldots, mn$,
let $S_j$ denote the copy of $S$ in $G_j$, and let $S' = K\cup \bigcup_{j = 1}
^{mn}S_j$. The set $S'\subseteq V(G')$ satisfies $|S'|= mn\gamma(G) + Bnm$.
Moreover, $S'$ is an $\alpha$-rate dominating set in $G'$:
\begin{itemize}
    \item For every $j= 1,\ldots, mn$ and for every $v\in V(G_j)$,
    the set $N_{G'}[v]\cap S'$ is the disjoint union of sets
    $N_{G_j}[v]\cap S_j$ and $N_{G'}(v)\cap K$.
Hence
$$|N_{G'}[v]\cap S'| =
|N_{G_j}[v]\cap S_j|+ |N_{G'}(v)\cap K|\ge$$$$\ge 1 + k_v\ge  \alpha(d_{G_j}(v)+1+k_v) = \alpha |N_{G'}[v]|\,.$$
        The second inequality holds by the choice of $k_v$.
    \item Let $v\in K$.
    By construction of $G'$, $v$ is adjacent to every other vertex in $K$, and to at most $n^2$ remaining vertices.
    Hence $|N_{G'}[v]|\le |K|+n^2=Bmn+n^2$. Moreover, $|N_{G'}[v]\cap S'|\ge |K| = Bmn$.
    Therefore, to show that  $|N_{G'}[v]\cap S'|\ge \alpha |N_{G'}[v]|$, it suffices to prove that
    $$\frac{Bmn}{Bmn+n^2}\ge \alpha\,.$$
    This is equivalent to
    $Bm(1-\alpha)\ge \alpha n\,,$
    or, equivalently,
    $$m\ge \frac{\alpha}{(1-\alpha)B}n\,,$$
    which is true by the definition of $m$ and since $n^{\epsilon/3} \ge 1\ge \frac{\alpha}{(1-\alpha)B}\,.$
\end{itemize}
This shows that $\gamma_{\times \alpha}(G') \le mn\gamma(G) + Bmn$
and completes the proof of the claim.

\medskip
%

Let us write $n' = |V(G')|$. By assumptions~(\ref{size-of-G-2'}) and (\ref{size-of-G-3'}) we have
$$n' = mn^2+Bmn \le
n^{3+2\epsilon/3} + Bn^{2+2\epsilon/3}
= (n+B)n^{2+2\epsilon/3}\le n^{1+\epsilon/3}n^{2+2\epsilon/3}= n^{3+\epsilon}\,.$$
Suppose that there exists a polynomial time algorithm $A'$ that computes a
$(\frac{1}{3}-\epsilon)\ln n'$-approximation to $\alpha$-rate domination in $G'$.
Let $S'$ be the set computed by $A'$. Then $|S'|\le (\frac{1}{3}-\epsilon)(\ln n') \gamma_{\times \alpha}(G')$.

Similarly as in the proof of the claim above, let $S'_j = S'\cap V(G_j)$
and pick an index $j^*\in \{1,\ldots, mn\}$ for which the value of $|S_j'|$ is smallest.
Then, setting $S=S'_{j^*}$ results in a dominating set in $G_j$ (and hence in $G$).

\begin{sloppypar}
We can bound the size of $S$ from above as follows:
$$
\begin{array}{rcll}
|S| &\le & \frac{1}{mn}\cdot|S'| & \textrm{(by the choice of $j^*$)}\\
& \le & \frac{1}{mn}(\frac{1}{3}-\epsilon)(\ln n') \cdot\gamma_{\times \alpha}(G')& \textrm{(by the assumption on $A'$)}\\
& \le & \frac{1}{mn}(\frac{1}{3}-\epsilon)(\ln(n^{3+\epsilon}))\cdot(mn\gamma(G) + Bnm)& \textrm{(by the Claim and $n' \le n^{3+\epsilon}$)}\\
& = & (\frac{1}{3}-\epsilon)({3+\epsilon})(\ln n)\cdot(\gamma(G) + B)& \\
& \le & (\frac{1}{3}-\epsilon)({3+\epsilon})(1+\epsilon)(\ln n)\cdot \gamma(G)& \textrm{(by (\ref{gamma'}))}\\
& = & (1-\epsilon')(\ln n)\gamma(G)\,, &
\end{array}$$
where $\epsilon' = \epsilon^3+11\epsilon^2/3+5\epsilon/3\in (0,1)$.
Therefore, $S$ approximates domination within a factor of
$(1-\epsilon')\ln n$. By Theorem~\ref{thm:D-inapprox}, this shows that there is no polynomial time algorithm approximating
$\alpha$-rate domination within a factor of \hbox{$(\frac{1}{3}-\epsilon)\ln n$}, unless
${\sf NP}\subseteq {\sf DTIME}(n^{O(\log\log n)})$.
\end{sloppypar}
\end{proof}


\begin{thebibliography}{99}
\bibitem{AMS06}
{\sc N.~Alon, D.~Moshkovitz} and {\sc S.~Safra}.
\newblock Algorithmic construction of sets for $k$-restrictions.
\newblock \emph{ACM Transactions on Algorithms} 2 (2006) 153--177.


\bibitem{CC04}
{\sc M.~Chleb\'ik} and {\sc J.~Chleb\'ikova}.
\newblock Approximation hardness of dominating set problems in bounded degree
graphs.
\newblock \emph{Information and Computation} 206 (2008) 1264--1275.


\bibitem{CMV-11}
{\sc F. Cicalese, M. Milani\v c} and {\sc U. Vaccaro}.
\newblock  Hardness,  approximability, and exact algorithms for vector domination and total vector domination in graphs.
\newblock {\em FCT 2011}, LNCS Vol. 6914 (2011) 288--297.

\bibitem{CMV12}
{\sc F.~Cicalese, M.~Milani\v c} and {\sc U.~Vaccaro},
\newblock On the approximability and exact algorithms for vector domination and related problems in graphs.
\newblock To appear in {\em Discrete Appl. Math.},
\url{http://dx.doi.org/10.1016/j.dam.2012.10.007}\,.

\bibitem{CLB81}
 {\sc D.G.~Corneil, H.~Lerchs} and {\sc L.~Stewart Burlingham}.
\newblock   Complement reducible graphs.
\newblock {\em Discrete Appl. Math.} 3 (1981) 163--174.

\bibitem{CPS85}
 {\sc D.G.~Corneil, Y.~Perl} and {\sc L.K.~Stewart}.
\newblock  A linear recognition algorithm for cographs.
\newblock {\em SIAM J.~Comput.} 14 (1985) 926--934.

\bibitem{DRV-2004}
{\sc F. Dahme, D. Rautenbach} and {\sc L. Volkmann}.
\newblock Some remarks on $\alpha$-domination.
\newblock {\em Discussiones Mathematicae, Graph Theory}
{24} (2004) 423--430.

\bibitem{DRV-2008}
{\sc F. Dahme, D. Rautenbach} and {\sc L. Volkmann}.
\newblock $\alpha$-Domination perfect trees.
\newblock {\em Discrete Math.} {308} (2008) 3187--3198.

\bibitem{DSNT}
{\sc  T.N. Dinh, Y. Shen, D.T. Nguyen} and
{\sc M.T. Thai}.
\newblock On the approximability of positive influence dominating set in social networks.
\newblock To appear in {\em J. Comb. Optim.}, 
\url{http://dx.doi.org/10.1007/s10878-012-9530-7}\,.

\bibitem{Dobson}
{\sc G. Dobson}.
\newblock Worst-case analysis of greedy heuristics for integer programming with nonnegative data.
\newblock {\em Mathematics of Operations Research}
7 (1982) 515--531.

\bibitem{DLN-11}
{\sc M.P. Dobson, V.~Leoni} and {\sc G. Nasin}.
\newblock The multiple domination and limited packing problems in graphs.
\newblock {\em Information Processing Letters} 111 (2011) 1108--1113.

\bibitem{DF-1997}
{\sc R. Duh} and {\sc M. F\"urer}.
\newblock Approximation of $k$-set cover by semi-local optimization.
\newblock In: Proceedings of the 29th ACM Symposium on Theory of Computing,
STOC, 1997, pp. 256--264.

\bibitem{DHLM-2000}
{\sc J.E. Dunbar, D.G. Hoffman, R.C. Laskar} and {\sc L.R. Markus}.
\newblock $\alpha$-Domination.
\newblock {\em Discrete Math.} {211} (2000) 11--26.


\bibitem{Feige-98}
{\sc U.~Feige.}
\newblock A threshold of $\ln n$ for approximating set cover.
\newblock \emph{Journal of ACM} 45 (1998) 634--652.

\bibitem{FJ-85a}
{\sc J.F. Fink} and {\sc M.S. Jacobson}.
\newblock $n$-domination in graphs.
\newblock {\em Graph Theory with Applications to Algorithms and Computer Science.}
John Wiley and Sons, New York, 1985, pp.~283--300.

\bibitem{FJ-85b}
{\sc J.F. Fink} and {\sc M.S. Jacobson}.
\newblock  On $n$-domination, $n$-dependence and forbidden subgraphs.
\newblock {\em Graph Theory with Applications to Algorithms and Computer Science.}
John Wiley and Sons, New York, 1985, pp.~301--311.

\bibitem{GPZ-09}
{\sc A. Gagarin, A. Poghosyan} and {\sc V.E. Zverovich}.
\newblock Upper bounds for $\alpha$-domination parameters.
\newblock {\em Graphs and Combinatorics} 25  (2009) 513--520.

\bibitem{GPZ-11}
{\sc A. Gagarin, A. Poghosyan} and {\sc V.E. Zverovich}.
\newblock Randomized algorithms and upper bounds for multiple domination in
graphs and networks.
\newblock {\em Discrete Applied Math.} (2011) doi:10.1016/j.dam.2011.07.004.

\bibitem{GJ79}
{\sc M.R.~Garey} and {\sc D.S.~Johnson},
Computers and Intractability: A Guide to the Theory of NP-Completeness, W. H. Freeman, 1979.

\bibitem{GH-07}
\textsc{W. Goddard} and {\sc M.A. Henning}.
\newblock Restricted domination parameters in graphs.
\newblock {\em Journal of Combinatorial Optimization} 13 (2007) 353--363.

\bibitem{HPV-99}
\textsc{J. Harant}, \textsc{A. Prochnewski} and \textsc{M. Voigt}.
\newblock{On dominating sets and independent sets of graphs}.
\newblock{\emph{Combinatorics, Probability and Computing}} 8 (1999) 547--553.

\bibitem{HH00}
\textsc{F. Harary} and \textsc{T.W. Haynes}.
\newblock{Double domination in graphs}.
\newblock{\emph{Ars Combin. }}
55 (2000) 201--213.


\bibitem{HHS1-98}
{\sc T.W.~Haynes, S.~Hedetniemi} and {\sc P.~Slater}
\newblock {\em Fundamentals of Domination in Graphs.}
\newblock Marcel Dekker, 1998.

\bibitem{HHS2-98}
{\sc T.W.~Haynes, S.~Hedetniemi} and {\sc P.~Slater} (Eds.)
\newblock {\em Domination in Graphs: Advanced Topics.}
\newblock Marcel Dekker, 1998.

\bibitem{HK-10}
{\sc M.A.~Henning} and  \textsc{A.P.~Kazemi}.
\newblock $k$-tuple total domination in graphs.
\newblock {\em Discrete Appl.~Math.} 158 (2010) 1006--1011.

\bibitem{JP-89}
\textsc{M. S. Jacobson} and  \textsc{K. Peters}.
\newblock \emph{Complexity questions for $n$-domination and related parameters.}
\newblock {\em Congr. Numer.} 68 (1989) 7--22.

\bibitem{KKT-05}
\textsc{D. Kempe, J.M. Kleinberg} and  \textsc{ E. Tardos}.
\newblock \emph{Influential Nodes in a Diffusion Model for Social Networks.}
ICALP 2005, 1127--1138.

\bibitem{KL04} \textsc{R.Klasing} and  \textsc{C. Laforest}.
\newblock  Hardness results and approximation algorithms of $k$-tuple domination in graphs.
\newblock \emph{Inform. Process. Lett}. 89 (2004) 75--83.

\bibitem{LC-02}
{\sc C.S.~Liao} and {\sc G.J.~Chang}.
\newblock Algorithmic aspects of $k$-tuple domination in graphs.
\newblock {\em Taiwanese Journal of Mathematics} 6 (2002) 415--420.

\bibitem{LC-03}
{\sc C.S.~Liao} and {\sc G.J.~Chang}.
\newblock $k$-tuple domination in graphs.
\newblock {\em Inform. Process. Lett.} 87 (2003) 45--50.

\bibitem{M-2012}
{\sc S. Mishra}.
\newblock Complexity of majority monopoly and signed domination problems.
\newblock{\em J. Discrete Algorithms} 10 (2012) 49--60.

\bibitem{MRS-2002}
{\sc S. Mishra, J. Radhakrishnan} and {\sc S. Sivasubramanian}.
\newblock On the hardness of approximating minimum monopoly problems.
\newblock{\em FST TCS 2002} LNCS Vol. 2556 2002, 277-288

\bibitem{MR-2006}
\textsc{\sc S. Mishra} and {\sc S. B.  Rao}.
\newblock Minimum monopolies in regular and tree graphs.
\newblock{\em Discrete Mathematics} 306 (2006) 1586--1594.

\bibitem{MR-07}
{\sc E.~Mossel}, and {\sc S.~Roch.}
\newblock  On the submodularity of influence in social networks.
\newblock {\em Proc. 39th Ann. ACM Symp. on Theory of Comp.}, ACM, 2007, pp.~128--134.



\bibitem{Peleg-02}
{\sc D.~Peleg.}
\newblock Local majorities, coalitions and monopolies in graphs: a review.
\newblock {\em Theoretical Computer Science} 282 (2002) 231--257.

\bibitem{RSS}
{\sc V. Raman, S. Saurabh} and {\sc S. Srihari}.
\newblock Parameterized Algorithms for Generalized Domination.
\newblock  {\em COCOA 2008},
LNCS Vol. 5165 (2008) 116--126.



\bibitem{Uehara-02}
{\sc R.~Uehara.}
\newblock Linear time algorithms on chordal bipartite and strongly chordal graphs
\newblock {\em Proc. ICALP 2002}, LNCS 2380, 993–-1004, 2002.

\bibitem{Wang11}
{\sc F. Wang, H. Du, E. Camacho, K. Xu, W. Lee, Y. Shi,} and {\sc S. Shan}
\newblock On positive influence dominating sets in social networks.
\newblock {\em Theoretical Computer Science} 412 (2011) 265--269.

\bibitem{Wolsey-82}
{\sc L.A.~Wolsey.}
\newblock An analysis of the greedy algorithm for the submodular set covering  problem.
\newblock {\em Combinatorica} 2 (1982) 385--393.


\bibitem{ZWZW-10}
{\sc F.~Zou, J. K Willson, Z.~Zhang}, and {\sc W.~Wu.}
\newblock Fast information propagation in social networks.
\newblock {\em Discrete Mathematics, Algorithms and Applications} 2 (2010) 1--17.
\end{thebibliography}
\end{document}